\documentclass[aps,prd,showpacs,floatfix,twocolumn,superscriptaddress,nofootinbib,preprintnumbers, 10pt]{revtex4-2} 

\makeatletter
\def\p@subsection{}
\makeatother

\usepackage{graphicx,amsmath,amssymb,lipsum,bm}

\usepackage[dvipsnames]{xcolor}
\definecolor{xlinkcolor}{rgb}{0.7752941176470588, 0.22078431372549023, 0.2262745098039215}

\usepackage{booktabs} 
\usepackage[normalem]{ulem}
\usepackage[utf8]{inputenc} 
\usepackage{mathtools}
\usepackage{aas_macros}
\usepackage{xspace}

\usepackage{ulem}
\usepackage{diagbox}

\usepackage[dvipsnames]{xcolor}
\usepackage{mathrsfs}

\usepackage[colorlinks=true,citecolor=xlinkcolor,linkcolor=xlinkcolor,urlcolor=xlinkcolor, backref=false,pdfborder={0 0 0}]{hyperref}

\newcommand{\be}{\begin{equation}}
\newcommand{\ee}{\end{equation}}
\newcommand{\beqa}{\begin{eqnarray}}
\newcommand{\eeqa}{\end{eqnarray}}

\renewcommand\k{{\bm k}}
\newcommand\q{\bm{q}}

\newcommand\GG{\Gamma_3}
\newcommand\G{\mathcal{G}_2}

\newcommand{\bseq}{\begin{subequations}}
\newcommand{\eseq}{\end{subequations}}

\renewcommand{\ln}{\mathop{\rm ln}\nolimits}

\def\ltsima{$\; \buildrel < \over \sim \;$\xspace}
\def\gtsima{$\; \buildrel > \over \sim \;$\xspace}
\def\simlt{\lower.5ex\hbox{\ltsima}}
\def\simgt{\lower.5ex\hbox{\gtsima}}

\newcommand{\hinvMpc}{\,h^{-1}\, {\rm Mpc}\,}
\newcommand{\Hunit}{\, {\rm km}\,s^{-1}\, {\rm Mpc}^{-1}\,}
\newcommand{\hMpcinv}{\,h\, {\rm Mpc}^{-1}\,}
\newcommand{\hinvGpc}{\,h^{-1}\, {\rm Gpc}\,}

\newcommand{\Mpch}{\, h^{-1}\mathrm{Mpc}\, }

\newcommand{\hkpc}{\, h\mathrm{kpc}^{-1}\, }
\newcommand{\kpch}{\, h^{-1}\mathrm{kpc}\, }
\newcommand{\kmax}{\, k_{\rm max}\, }
\newcommand{\knl}{\, k_{\rm NL}\, }

\newcommand{\Lya}{Ly-$\alpha$\xspace}
\newcommand{\td}{\delta}
\newcommand{\dd}{\partial}

\newcommand{\pvec}{\mathbf{p}}
\newcommand{\kvec}{\mathbf{k}}

\newcommand{\kpar}{k_{\parallel}}

\newcommand{\kvperp}{\mathbf{k}_{\perp}}
\newcommand{\apar}{\alpha_{\parallel}}
\newcommand{\aperp}{\alpha_{\perp}}
\newcommand{\alphaiso}{\alpha_{\mathrm{iso}}}
\newcommand{\alphaap}{\alpha_{\mathrm{ap}}}

\def\gsim{\raise0.3ex\hbox{$\;>$\kern-0.75em\raise-1.1ex\hbox{$\sim\;$}}}
\def\lsim{\raise0.3ex\hbox{$\;<$\kern-0.75em\raise-1.1ex\hbox{$\sim\;$}}}

\def\beqn#1{\begin{equation}\label{#1}}
\def\eeqn{\end{equation}}

\def\beqa#1{\begin{eqnarray}\label{#1}}
\def\eeqa{\end{eqnarray}}

\def\kmax{{k_\text{max}}}

\def\Mpch{h^{-1}{\text{Mpc}}}

\def\Z2{$\mathcal{Z_2}$}

\newcommand {\ignore}[1]{}

\renewcommand{\arraystretch}{1.3} 
\begin{document}

\preprint{MIT-CTP/5720}

\title{The ACCEL$^2$ Project: Precision Measurements of EFT Parameters and BAO Peak Shifts for the Lyman-$\alpha$ Forest}

\author{Roger de Belsunce}
\email{rbelsunce@berkeley.edu}
\affiliation{Lawrence Berkeley National Laboratory, One Cyclotron Road, Berkeley CA 94720, USA}
\affiliation{Berkeley Center for Cosmological Physics, Department of Physics, UC Berkeley, CA 94720, USA}
\author{Shi-Fan Chen}
\email{sfschen@ias.edu}
\affiliation{Institute for Advanced Study, 1 Einstein Drive, Princeton, NJ
08540, USA}
\author{Mikhail M. Ivanov}
\email{ivanov99@mit.edu}
\affiliation{Center for Theoretical Physics, Massachusetts Institute of Technology, 
Cambridge, MA 02139, USA}
 \affiliation{The NSF AI Institute for Artificial Intelligence and Fundamental Interactions, Cambridge, MA 02139, USA}
\author{\\ Corentin Ravoux}\affiliation{Universit\'{e} Clermont-Auvergne, CNRS, LPCA, 63000 Clermont-Ferrand, France}
\author{Sol{\` e}ne Chabanier}\affiliation{Lawrence Berkeley National Laboratory, One Cyclotron Road, Berkeley CA 94720, USA}
\author{Jean Sexton}\affiliation{Lawrence Berkeley National Laboratory, One Cyclotron Road, Berkeley CA 94720, USA}
\author{Zarija Luki{\' c}}\affiliation{Lawrence Berkeley National Laboratory, One Cyclotron Road, Berkeley CA 94720, USA}

\begin{abstract} 
We present precision measurements of the bias parameters of the one-loop power spectrum model of the Lyman-$\alpha$ (\Lya) forest, derived within the effective field theory of large-scale structure (EFT). We fit our model to the three-dimensional flux power spectrum measured from the ACCEL$^2$ hydrodynamic simulations. The EFT model fits the data with an accuracy of below 2\% up to $k = 2 \hMpcinv$. Further, we analytically derive how non-linearities in the three-dimensional clustering of the \Lya forest introduce biases in measurements of the Baryon Acoustic Oscillations (BAO) scaling parameters in radial and transverse directions. From our EFT parameter measurements, we obtain a theoretical error budget of $\Delta \apar = -0.2\%$  ($\Delta \aperp = -0.3\%$) for the radial (transverse) parameters at redshift $z=2.0$. This corresponds to a shift of $-0.3\%$ ($0.1\%$) for the isotropic (anisotropic) distance measurements. We provide an estimate for the shift of the BAO peak for \Lya\ -- quasar cross-correlation measurements assuming analytical and simulation-based scaling relations for the non-linear quasar bias parameters resulting in a shift of $-0.2\%$ ($-0.1\%$) for the radial (transverse) dilation parameters, respectively. This analysis emphasizes the robustness of \Lya forest BAO measurements to the theory modeling. We provide informative priors and an error budget for measuring the BAO feature -- a key science driver of the currently observing Dark Energy Spectroscopic Instrument (DESI). Our work paves the way for full-shape cosmological analyses of \Lya forest data from DESI and upcoming surveys such as the Prime Focus Spectrograph, WEAVE-QSO, and 4MOST. 
\end{abstract}

\maketitle

\section{Introduction}
Neutral hydrogen in the low-density, highly ionized intergalactic medium (IGM) absorbs light, creating a series of distinctive absorption features in quasar spectra known as the Lyman-$\alpha$ (\Lya) forest. These absorption patterns trace the distribution of neutral hydrogen along the line of sight down to very small scales ($k\sim 5-10 \hMpcinv$). In this regime, measurements of the line-of-sight power spectrum~\cite{McDonald06, PYB13, DESI:2023xwh, Karacayli:2023afs} yield information on neutrino masses \citep{Seljak:2005, Viel:2010, PYB13, Palanque2020}, primordial black holes \citep{Afshordi:2003, Murgia:2019}, (warm) dark matter models \citep{Viel:2013, Baur:2016, Irsic17, Kobayashi:2017, Armengaud:2017, Murgia:2018,Garzilli:2019, Irsic:2020, Rogers:2022, Villasenor:2023, Irsic:2023}, early dark energy models \citep{2023PhRvL.131t1001G}, and thermal properties of the ionized (cold) IGM \citep{Zaldarriaga:2002, Meiksin:2009,McQuinn:2016, Viel:2006, Walther:2019, Bolton:2008, Garzilli:2012, Gaikwad:2019, Boera:2019, Gaikwad:2021, Wilson:2022, Villasenor:2022}. The three-dimensional clustering of the \Lya forest sets increasingly tight constraints on the expansion history of the Universe through the measurement of baryonic acoustic oscillations \citep[BAO;][]{McDonald:2007,Slosar2013, Busca:2013, duMasdesBourboux:2020pck, DESI_lya_2024} in the matter-dominated regime ($2\leq z \leq 4$). 

\Lya surveys have dramatically improved in accuracy, size, and depth, resulting in large samples of medium resolution spectra from the extended Baryon Oscillation Spectroscopic Survey (eBOSS;~\cite{eBOSS:2015jyv}) and the currently observing Dark Energy Spectroscopic Instrument (DESI;~\cite{DESI:2016fyo}). DESI will observe over its lifetime approximately $840,000$ \Lya forest skewers at $z>2.1$~\cite{DESI:2022}. Advances in theoretical modeling through the development of the one-loop power spectrum of the \Lya forest in the framework of the effective field theory of large-scale structure (EFT)\footnote{The EFT is a perturbative program describing large-scale dynamics only using the relevant symmetries to the tracer~\cite{Baumann:2010tm,Carrasco:2013mua,Ivanov:2022mrd}, \textit{i.e.}~for the \Lya forest the natural symmetries are the equivalence principle and rotations around the line of sight $\hat{z}$, denoted by the $SO(2)$ group, see, e.g.,~\cite{McDonald:2009dh,Givans:2020sez,Desjacques:2018pfv,Chen:2021rnb,Ivanov2024,Ivanov:2024jtl}.}, paired with beyond-BAO measurements using redshift space distortions (RSD) and the Alcock-Paczynski (AP) effect in configuration space~\cite{Gerardi:2022ncj, Cuceu:2021hlk}, and the measurement of the three-dimensional power spectrum~\cite{Font-Ribera:2018, Abdul-Karim:2023sgf, deBelsunce:2024knf, Horowitz:2024nny}, pave the way to extract more cosmological information from the \Lya forest by, e.g., directly fitting cosmological parameters~\cite{Gerardi:2022ncj}.\footnote{This approach is analogous to  EFT-based full-shape analyses
of the galaxy clustering data~\cite{Ivanov:2019pdj,DAmico:2019fhj,Chen:2021wdi,Philcox:2021kcw,Chen:2022jzq,Chen:2024vuf}.} 

In this work, we present a precision measurement of the EFT bias parameters from ACCEL$^2$ hydrodynamical simulations~\cite{Chabanier:2024knr}. To calibrate the large number of EFT bias parameters we require simulations with large enough box sizes to have enough quasilinear modes to study the deviations from linearity paired with sufficient resolution in order to fully resolve the \Lya forest, both of which are provided by the ACCEL$^2$ simulations. Our baseline results are based on the box of size $L=160 \hinvMpc$ with a resolution down to $25 \hkpc$ and we use one large box of size $L=640 \hinvMpc$ with a resolution down to $100 \hkpc$. The high redshift of the \Lya forest gives access to more linear modes than with, e.g., galaxy surveys, effectively ``pushing'' the non-linear scale to higher wavenumbers $k \sim 3-5 \hMpcinv$. We demonstrate that the one-loop \Lya power spectrum description provides a good fit at the $10\%$-level on large ($k\simlt 0.4 \hMpcinv$) scales and at the $2\%$-level on small ($k\simgt 1 \hMpcinv$)  scales to the three-dimensional power spectrum for the small box. Similarly for the large simulation box, we obtain residuals at the $7\%$ ($1\%$) level on large (small) scales, respectively. Previous analyses have performed similar fits using phenomenological fitting functions obtained from hydrodynamical simulations with sub-5\% accuracy down to $k\sim 10 \hMpcinv$~\cite{McDonald:2001fe, Arinyo-i-Prats:2015vqa, Givans2022}.  

A key science driver of the currently observing DESI Stage-IV spectroscopic survey~\cite{DESI:2016fyo} is to measure the expansion of the Universe through the measurement of baryon acoustic oscillations (BAO;~\cite{DESI_BAO_2024, DESI_lya_2024}). The BAO has become a pillar of cosmological analysis since its detection more than two decades ago and has been used to place tight constraints on its constituents and theories of gravity~\cite{SDSS:2005xqv, Cole:2005sx}. The BAO feature originates from oscillations in the baryon-photon fluid prior to decoupling which left an imprint at a characteristic length scale of $\sim 147\, \mathrm{Mpc}$ in both the cosmic microwave background (CMB;~\cite{Planck2018}) and the matter density. This measurement provides a ``standard ruler'' based on the distance, $r_d$, a sound wave can travel prior to the baryon-drag epoch, post recombination. DESI has provided high-precision measurements of the BAO feature in seven redshift bins from over 6 million extragalactic objects starting at $z=0.1$ to the three-dimensional \Lya forest clustering yielding the highest-redshift measurement at $z=2.33$~\cite{DESI_BAO_2024, DESI_lya_2024}. Connecting the BAO scale to constraints from, e.g.,~the CMB  or big-bang nucleosynthesis yields an ``absolute'' distance measurement. Whilst the BAO feature is remarkably robust, non-linearities in clustering modify it in two significant ways (see, e.g.,~\cite{Chen:2024tfp}): First, the amplitude of the oscillatory linear signal is damped towards small scales by incoherent bulk motions on the BAO scale. Second, out-of-phase contributions from clustering induce a shift in the observed BAO position. DESI is forecasted to reach a cumulative precision below 0.2\% over its five-year survey combining all tracers and redshifts~\cite{DESI:2016fyo}, requiring exquisite knowledge over theoretical and modeling systematics. 

In this work, we derive an error budget to account for theoretical and modeling systematics when fitting the BAO using \Lya forest data. We extend the approach presented in the context of the well-understood galaxy clustering~\cite{Chen:2024tfp} to the \Lya forest. Whilst the \Lya forest has made remarkably precise BAO measurements, theoretical uncertainties are much less quantified.\footnote{Measurements of compressed statistics (e.g., BAO, RSD, or AP parameters) require an estimate of the theoretical and systematic error budget in contrast to full-shape analyses which include and, crucially, propagate this uncertainty into the final parameter constraints~\cite{Chen:2024tfp}.} Recent simulation-based measurements of the BAO feature using a large suite of \Lya forest simulations ($N=1,000$) with a volume of $V=1\hinvGpc$ at redshift two, for example, suggest a large negative shift in the BAO peak~\cite{Sinigaglia:2024kub} 
at the $\sim -1\%$ ($\sim -0.3\%$)-level in redshift (real) space, respectively. An enormous effort is underway in the community to provide high-fidelity simulations that capture large-enough volumes for cosmological analyses whilst providing sufficient resolution for the small-scale behavior of the \Lya forest~\cite{Farr20, Sinigaglia:2024kub, Chabanier:2024knr}. 

In this paper, we use the ACCEL$^2$ hydrodynamic simulations to set constraints on the EFT bias parameters of the one-loop \Lya forest spectrum. We then use the parameter combinations explored by the Markov Chain Monte Carlo chains to obtain an estimate (and, crucially, an uncertainty) for the BAO shift. This analysis provides a theoretical and systematic error budget when fitting BAO measurements of the \Lya forest~\cite{duMasdesBourboux:2020pck, DESI_lya_2024}. Since most of the constraining power of the \Lya forest is driven by the cross-correlation with quasars, we also provide an estimate of the BAO shift in the cross-correlation~\cite{Font-Ribera:2013fha, duMasdesBourboux:2020pck, DESI_lya_2024}. As the ACCEL$^2$ simulations do not include a quasar (or halo) catalog, we use best-fit EFT parameters for higher-order quasar biases
from the eBOSS survey~\cite{Chudaykin:2022nru}, bias relations measured from clustering of halos in N-body simulations~\cite{Abidi:2018eyd}, and analytic bias relations to characterize nonlinearities in quasar clustering. This work presents the first step towards a full-shape analysis of the \Lya forest and quasar clustering in the EFT framework.  

This paper is organized as follows: in Sec.~\ref{sec:simulations} we present the employed ACCEL$^2$ hydrodynamical simulations of the \Lya forest. In Sec.~\ref{sec:LyaEFT} we briefly summarize our theoretical model based on the one-loop \Lya forest power spectrum in the EFT framework. We present our Gaussian likelihood and discuss details of the fitting procedure in   Sec.~\ref{sec:logL}. In Sec.~\ref{sec:BAO_shift} we derive an analytic formula for the non-linear shift of the BAO feature in redshift space. In Sec.~\ref{sec:results} we present our results and conclude in Sec.~\ref{sec:conclusion}. 

\section{ACCEL$^2$ simulations} \label{sec:simulations}
In the present work, we perform precision measurements of the EFT bias parameters on a suite of ACCEL$^2$ (ACCELerated expansion of the universe with ACCELerated computing on GPUs) hydrodynamical simulations of the \Lya forest~\cite{Chabanier:2024knr}. We use the ACCEL$^2$ simulation with the highest resolution with a box of length $L=160 \Mpch$ with $N=6144^3$ particles, yielding an effective resolution of $25 \kpch$, denoted by \texttt{ACCL2\_L160R25}, for our baseline results. To access more linear modes, we compare our results to the largest available box of size $L=640 \hinvMpc$ with an effective resolution of $100\, h^{-1}\mathrm{kpc}$, denoted by \texttt{ACCL2\_L640R100}. The input cosmology is based on the best fit $\Lambda$CDM-parameters obtained by the \textit{Planck} satellite $\Omega_b=0.0487,\, \Omega_m=0.31,\, H_0=67.5\Hunit,\, n_s=0.96$ and $\sigma_8=0.83$~\cite{Planck:2015fie}. Each simulation is initialized at $z=200$ and we use five snapshots at $z=2.0,\,2.6,\,3.0,\,3.6,\,4.0$. 

The simulations are based on the simulation code \texttt{Nyx}\footnote{Publicly available at \url{https://amrex-astro.github.io/Nyx/}.}. Briefly, \texttt{Nyx} is a highly scalable, adaptive mesh refinement (AMR) cosmological simulation code designed for modeling the evolution of baryonic gas and dark matter in an expanding universe~\cite{Almgren:2013, Sexton2021}. It solves the equations of compressible hydrodynamics for baryons, using an N-body approach for dark matter. The hydrodynamics implementation is formulated in Eulerian coordinates. Self-gravity, incorporating both gas and dark matter, is addressed by solving the Poisson equation through a geometric multi-grid method.  For details on how the \Lya forest is modeled, we refer the reader to~\cite{Lukic:2015} for a fuller presentation. In summary, the \Lya forest is modeled by tracking the abundance of six species: neutral and ionized hydrogen, neutral, singly and doubly ionized helium, and free electrons. For each species, the relevant atomic processes are simulated which includes ionization, recombination, and free-free transitions. Ref.~\cite{Chabanier:2024knr} provides evidence for resolution convergence of the used simulations. 

In the present work, we use flux fluctuations \be \label{eq:deltaF} \td_F = \frac{F}{\overline{F}(z)}-1\,,\ee around the mean value of transmission $\overline{F}(z)$ at some redshift $z$. The transmitted flux fraction is defined as  $F(x)=\exp{(-\tau(x))}$ where $\tau$ is the optical depth. The quantity of interest in the present work is the three-dimensional flux power spectrum, $P_{\rm 3D}(k,\mu)$ (or $P(k,\mu)$ in the following), which is the average of the squared norm of the Fourier transform of Eq.~\eqref{eq:deltaF} in bins of the Fourier wavenumber $k$ with the cosine of the angle to the line-of-sight, $\mu=\kpar/k$. We fit our theoretical model, introduced in the following section, to the measured $P(k,\mu)$. The power spectra are computed over each of the three axes and averaged together~\cite{Chabanier:2024knr}. We use a linear spacing in $k$ of $\Delta k \approx 0.039 \hMpcinv$ and in $\mu$-bins of $\Delta \mu = 0.25$.\footnote{To account for mode discreteness, we use the effective $k$ and $\mu$ for each spectrum. Note that our k bins are quite narrow, so we do not expect the binning
corrections to affect our results, see~\cite{Nishimichi:2020tvu,Philcox:2021kcw}
for related discussions in the context of galaxy clustering data.} We use a diagonal Gaussian covariance based on the number of expected Fourier modes per bin $P(k,\mu)\sqrt{2/N(k,\mu)}$, assuming that the measured power along each of the axes is independent. Whilst the measured power spectra are provided up to $k\sim 10^2 \hMpcinv$, we will use $\kmax = 0.5 - 5 \hMpcinv$ for the presented fits. 

\section{Theoretical Model} \label{sec:LyaEFT}
One of the key advantages of the \Lya forest is that it probes the Universe at intermediate redshifts ($2 \leq z \leq 4$), with access to many more linear modes than, e.g., galaxy surveys. In the following, we briefly summarize the theoretical model of the \Lya forest one-loop (auto) power spectrum in the EFT framework~\cite{Ivanov2024}. The model consists of four components
\begin{equation} \label{eq:Pmodel}
    P^{\rm th.}(k) = P^{\rm tree}(k) + P^{\rm 1-loop}(k) + P^{\rm ct}(k) + P^{\rm st.}(k) \,,
\end{equation}
where $k$ is the wavenumber, $P^{\rm tree}(k)$ the Kaiser power spectrum, $P^{\rm 1-loop}(k)$ the one-loop \Lya power spectrum, and $P^{\rm ct}(k)$ and $P^{\rm st.}(k)$ are the counter terms and stochastic contributions, respectively.\footnote{These counter terms can be understood as marginalizing over two-loop contributions. Physically, ``short-range non-locality'' can be described on scales larger than the non-locality scale $R$ through the addition of bias terms proportional to powers of  $k^2 R^2$ with order unity  coefficients~\cite{2009JCAP...08..020M}.} In the following we introduce each component separately. In addition, we resum the expression in Eq.~\eqref{eq:Pmodel} to include the non-perturbative action, i.e. damping, of long-wavelength (IR) displacements on baryon acoustic oscillations (BAO) in the linear power spectrum.\footnote{To control long-wavelength displacements, we perform infrared (IR) resummation for the redshift-space power spectrum, similar to the approach derived within the time-sliced perturbation theory~\cite{Blas:2015qsi,Blas:2016sfa,Ivanov:2018gjr,Vasudevan:2019ewf} and other approaches \cite{Vlah16,Chen24}.}  Note that 
all quantities that appear
in eq.~\eqref{eq:Pmodel}
are evaluated at an
effective redshift of 
the forest $z$--in this case the redshift of the simulation snapshot---and in what follows
we suppress the explicit
time-dependence. 

In linear theory, the tidal field is related to the dimensionless gradient of the peculiar velocity, $v_z$, along the line-of-sight (for a coordinate along the line-of-sight $x_z$) $\eta = -\dd_{x_z}v_z/\mathcal{H}$ through~\cite{McDonald:1999dt, Arinyo-i-Prats:2015vqa} \be \label{eq:delta_Kaiser} \td_F = b_1\td + b_{\eta}\eta \,, \ee where $\mathcal{H}=aH$ is the conformal Hubble parameter given by the scale factor $a$ and the Hubble parameter $H$. The well-known Kaiser formula connects the redshift-space flux power to the linear matter power spectrum through~\cite{Kaiser:1987qv, McDonald:1999dt, McDonald:2001fe} 
\be \label{eq:IR-Kaiser}P^{\rm tree}(k,\mu) = K^2_1(\k) P_{\rm lin}(k)\,,\,\,\, K_1(\k)\equiv (b_1-b_{\eta}f\mu^2)\,,
\ee 
where $\mu$ is the angle of $k=\{\kpar,\kvperp\}$ to the line-of-sight, $\mu \equiv \kpar/k$, and $f$ the (linear) growth rate.\footnote{Note that our convention absorbs the negative sign into $b_{\eta}$~\cite{Ivanov2024}.}

The one-loop contribution is given by~\cite{Ivanov2024}
\begin{align}
P^{\rm 1-loop}& (k,\mu) 
=2\int_{\q} K_2^2(\q,\k-\q)
P_{\text{lin}}(|\k-\q|)P_{\text{lin}}(q)  \nonumber\\
&\quad + 6 K_1(\k)P_{\text{lin}}(k)\int_{\q} K_3(\k,-\q,\q)P_{\text{lin}}(q)\,.
\end{align}
The redshift-space kernels, $K_{2,3}$, are given in Eqs.~(3.19)  in Ref.~\cite{Ivanov2024} and the counter terms are given by
\begin{align}
\label{eq:shoch}
P^{\rm ct}(k,\mu) =
&-2(c_0+c_2\mu^2+c_4\mu^4)K_1(\k)k^2 P_{\text{lin}}(k)\,,
\end{align}
where the angular dependence 
is regulated by $c_0,c_2,c_4$.
The stochastic contributions are given by 
\begin{align}
P^{\rm st.}(k,\mu) =
&P_{\text{shot}}+a_0\frac{k^2}{\knl^2}+a_2\frac{k^2\mu^2}{\knl^2}\,,
\end{align}
where $P_{\text{shot}}$ is a constant shot noise term which can, e.g., be estimated from the one-dimensional \Lya power spectrum. The parameters $a_0$ and $a_2$ are Wilson coefficients that cancel the UV sensitivity in the integrals in Eq.~(6.4) in Ref.~\cite{Ivanov2024}.

Note that the physically observed stochastic 
contribution is the sum of the 
power spectrum of the stochastic field
and the cutoff-sensitive part 
of the loop integral, which introduces 
an error in our loop calculations. 
Schematically, using the auto-spectrum
of the $\delta^2$ field as a proxy
for all constant deterministic 
contributions, we have:
\be 
\label{eq:pepsphys}
\begin{split}
& P_{\rm shot}^{\rm phys.} =P_{\epsilon\epsilon}+ P_{\rm shot}^{\rm ctr.} + b_2^2 \mathcal{I}_{\delta^2\delta^2}^{\Lambda}~\,,\\
& \mathcal{I}_{\delta^2\delta^2}^{\Lambda}(z)=\int_{\Lambda}^\infty \frac{dp}{4\pi^2}~p^2P^2_{11}(p,z) \,,
\end{split}
\ee 
where $\Lambda\sim k_{\rm NL}$
is a UV cutoff.
For galaxies $P_{\epsilon\epsilon} \sim \bar n^{-1}\sim 5\cdot 10^3~[\Mpch]^3$ (a number for the BOSS CMASS sample~\cite{BOSS:2016wmc}), which is much greater 
than the above rightmost contribution over the 
small scale modes, e.g. for $z=0.5$ we have $\mathcal{I}_{\delta^2\delta^2}^{\Lambda}\simeq 50$ $~[\Mpch]^3$.
For the \Lya forest, 
thanks to a 
huge number of absorbing atoms, 
$P_{\epsilon\epsilon}$ is virtually zero,\footnote{We ignore
effects of the
exclusion of \Lya
absorbers, 
see, e.g.,~\cite{Irsic:2018hhg}
for a discussion
in the context of 
1D correlations.} 
while the 
cutoff loop 
part is 
\[
\mathcal{I}_{\delta^2\delta^2}^{\Lambda}(z=3)\simeq 0.03~[\Mpch]^3\sim \mathcal{O}(1)\times \frac{1}{k_{\rm NL}^3}\,.
\]
The naturalness
arguments 
imply that 
$P_{\rm shot}^{\rm phys.}$
cannot be smaller
than individual 
terms in the r.h.s. of Eq.~\eqref{eq:pepsphys}. 
Hence, using the
largest of them, we get an estimate:
\be 
P_{\rm shot}^{\rm phys.}(z=3)\sim \frac{1}{k_{\rm NL}^3}\,.
\ee 
In a power-law universe, $P_{\rm lin} \propto k^n$, the relevance of 
$P_{\rm shot}^{\rm phys.}$ 
depends on the power index. For $n\approx -2$ it is roughly of the same order as the 
two-loop corrections
and counterterms,
and naively can be 
ignored. The 
scaling Universe 
estimate, however,
may not be very adequate 
given the high precision
of the ACCEL$^2$ data.
Given this reason,
in this work we 
consider two types
of analysis: 
the minimal one, 
where we ignore 
$P_{\rm shot}^{\rm phys.}$
and the $k^2$ -- 
counterterms,
and the extended
one where we include them. 
In addition, in the latter analysis, we also
include the scale-dependent
shot noise corrections
given by Eq.~\eqref{eq:shoch}.
Jumping ahead, let us say that 
the high redshift
data are consistent
with $P_{\rm shot}=a_0=a_2=0$,
in agreement
with the 
scaling Universe estimate,
while the low-redshift data
do prefer some non-zero stochastic terms
to compensate for the error in
the one-loop 
calculations. This can be 
understood through the change
of the effective scaling 
index in a power law universe,
which makes the stochastic 
corrections more important.

The full model for the one-loop \Lya power spectrum is described by the following set of nuisance parameters\footnote{Following~\cite{Ivanov2024}, we set $b_{\GG}=0$ due to the degeneracy with $b_{\G}$. Constraints could be obtained by including cross-correlation with matter, halos or quasars (see, e.g.,~\cite{Arinyo-i-Prats:2015vqa,Givans2022}). We leave this to future work.} \be \label{eq:nuissance_param} \{b_1, b_\eta, b_2, b_{\mathcal{G}_2}, b_{(KK)_\parallel}, b_{\Pi^{[2]}_\parallel}, b_{\delta \eta}, b_{\eta^2}\}\,,\ee in addition to the cubic EFT terms \be \label{eq:cubic_param} \{b_{\Pi^{[3]}_\parallel},b_{(K\Pi^{[2]})_\parallel},b_{\delta\Pi^{[2]}_\parallel},b_{\eta\Pi^{[2]}_\parallel}\}\,,\ee and the counter- and stochastic terms \be \label{eq:ct_param}  \{c_{0,2,4},a_{0,2},P_{\rm shot}\}\,,\ee yielding a total of 18 parameters.

\section{Likelihood} \label{sec:logL}
We calibrate the EFT bias parameters for the one-loop \Lya power spectrum by fitting a model to the power spectrum measured from ACCEL$^2$ hydrodynamical simulations~\cite{Chabanier:2024knr}. Therefore, we follow the procedure outlined in Ref.~\cite{Ivanov2024,Givans2022} and fit the power spectrum by sampling the $\chi^2$ function
\begin{equation} \label{eq:chi2}
    \chi^2 = \sum_i \frac{\left[P_i^{\rm data}-P^{\rm model}(k_i,\mu_i)\right]^2}{2 \left(P_i^{\rm data}\right)^2/N_i}\,,
\end{equation}
where $P_i^{\rm data}$ is the data vector measured from the ACCEL$^2$ simulations in bins of wavenumber $k$ and angle $\mu = \kpar/k$ with $N_i$ Fourier modes per bin and $P^{\rm model}(k_i,\mu_i)$ is the theory prediction for the \Lya forest EFT power spectrum, given in Eq.~\eqref{eq:Pmodel}. To account for mode discreteness, we evaluate the model vector at the power spectrum-weighted mean wavenumber $k$ and angle $\mu$ for each bin, computed via $\overline{x}=\sum_i P_i^{\rm data}x_i/\sum_i P_i^{\rm data}$ for $x\in \{k,\mu\}$. We jointly fit all four $\mu$ bins. Note that we do not introduce a noise floor, as done in Ref.~\cite{Givans2022, Chabanier:2024knr}, for the key results of the paper. This has the advantage that the full sensitivity of the simulations can be used to constrain the bias parameters.\footnote{We obtain consistent results when introducing a noise floor, effectively down-weighting small scales, with larger uncertainties on the recovered parameters, as expected.}

To reduce the computational burden of sampling the EFT parameters, we analytically marginalize over parameters that enter the theoretical model linearly such as the cubic EFT terms and the counter terms given in Eqs.~\eqref{eq:cubic_param} and \eqref{eq:ct_param}, respectively. The theory prediction can be split into theory templates $\mathbf{A}$ and $\mathbf{B}$ that depend on a set of model parameters $\mathbf{\theta}$ and a linear dependence on parameters $\mathbf{\phi}_i$ (over which we analytically marginalize) $\mathbf{t}(\mathbf{\theta},\phi)\equiv \mathbf{A}(\mathbf{\theta}) + \sum_i\phi_i\mathbf{B}_i(\mathbf{\theta})$. This assumes that the likelihood and priors on linear parameters are Gaussian (see, e.g.,~\cite{Philcox:2020zyp,Sailer:2024coh}). In practice, this means that we do not sample the cubic EFT parameters, counterterms or stochastic terms
\be 
\{b_{\Pi^{[3]}_\parallel},b_{(K\Pi^{[2]})_\parallel},b_{\delta\Pi^{[2]}_\parallel},b_{\eta\Pi^{[2]}_\parallel}, c_{0,2,4}, P_{\rm shot},a_{0,2}\}\,,
\ee 
which drastically reduces the number of sampled parameters from 15 (or 18 when including stochastic terms) down to eight. The sampled (marginalized) parameters are shown in the left (right) column of Tab.~\ref{tab:params_priors}.  For the present analysis, we fix the cosmological parameters of our EFT model to the input values of the \texttt{Nyx} simulations~\cite{Chabanier:2024knr}, described in Sec.~\ref{sec:simulations}. 

\begin{table}
\centering
\begin{tabular}{cc|cc}
\hline
\hline
EFT Parameter  & Prior & EFT Parameter & Prior \\
(Sampled)      & &  (Marginalized) &  \\
\hline
$b_1$ & $\mathcal{U}(-2, 2)$ & $\frac{c_{0,2,4,6}}{[h^{-1}\text{Mpc}]^2}$ & $\mathcal{N}(0, 1^2)$ \\
$b_{\eta}$ & $\mathcal{U}(-2, 2)$ & $P_{\text{shot}}$ & $\mathcal{N}(0, 5^2)$ \\
$b_2$ & $\mathcal{N}(0, 2^2)$ & $\frac{a_{0,2}}{[h^{-1}\text{Mpc}]^2}$ & $\mathcal{N}(0, 5^2)$ \\
$b_{\mathcal{G}_2}$ & $\mathcal{N}(0, 2^2)$ & $b_{(K\Pi^{[2]})_\parallel}$ & $\mathcal{N}(0, 2^2)$ \\
$b_{(KK)_\parallel}$ & $\mathcal{N}(0, 2^2)$ & $b_{\delta\Pi^{[2]}_\parallel}$ & $\mathcal{N}(0, 2^2)$ \\
$b_{\Pi^{[2]}_\parallel}$ & $\mathcal{N}(0, 2^2)$ & $b_{\eta\Pi^{[2]}_\parallel}$ & $\mathcal{N}(0, 2^2)$ \\
$b_{\delta\eta}$ & $\mathcal{N}(0, 2^2)$ & $b_{\Pi^{[3]}_\parallel}$ & $\mathcal{N}(0, 2^2)$ \\
$b_{\eta^2}$ & $\mathcal{N}(0, 2^2)$ & & \\
\hline
\end{tabular}
\caption{Tabulated EFT parameters for the one-loop \Lya forest power spectrum with corresponding priors used in the present analysis. Uniform priors with lower $x_1$ and upper bound $x_2$ are denoted by $\mathcal{U}(x_1, x_2)$ and Gaussian priors with mean $\mu$ and standard deviation $\sigma$ are quoted as $\mathcal{N}(\mu, \sigma)$.}
\label{tab:params_priors}
\end{table}

We explore the EFT parameter space using a Markov Chain Monte Carlo (MCMC) technique, implemented within the \texttt{Montepython} sampler~\cite{Brinckmann:2018cvx, Audren:2011ne}. The chains are considered converged when the Gelman-Rubin diagnostic satisfies $R -1 < 0.01$ for all parameters~\cite{1992StaSc...7..457G}. Typically, this requires $\sim 0.1$ CPU-hours (using one AMD Milan CPU on the Perlmutter computer at NERSC). We analyze and plot the 1D and 2D marginalized posteriors with \texttt{getdist}~\cite{Lewis:2019xzd}. The linear power spectrum, $P_{\rm lin}(k)$, 
and the decomposition into an oscillatory (``wiggly'') and a smooth (``non-wiggly'')
part, required for IR 
resummation, 
are generated with the Boltzmann solver \texttt{CLASS-PT}~\cite{Diego_Blas_2011,Chudaykin:2020aoj}. 

\section{Non-linear BAO shift} \label{sec:BAO_shift}
Measuring the BAO feature in the distribution of galaxies and 3D clustering of the \Lya forest is one of the most robust and accurate ways to measure the expansion history of our Universe. Since its detection in spectroscopic surveys by BOSS~\cite{SDSS-BAO} and the 2dF Galaxy Redshift Survey~\cite{Cole:2005sx} and subsequently in the 3D clustering of the \Lya forest~\cite{Busca:2013, Slosar2013}, it has become a pillar of precision cosmology to measure the expansion of the Universe~\cite{DESI_BAO_2024, DESI_lya_2024}. The standard approach taken in BAO analyses to measure the expansion history of our Universe, is to compress these rich (tomographic) data sets into two numbers corresponding to the radial ($\apar$) and transverse ($\aperp$) BAO distance scales. Briefly, the BAO parameters are defined as
\begin{equation}
    \apar \equiv \frac{H^{\rm fid}(z)r_d^{\rm tem}}{H(z)r_d}\, \qquad \aperp \equiv \frac{D_A(z)r_d^{\rm tem}}{D_A^{\rm fid}(z)r_d}\,,
\end{equation}
where $H(z)$ and $D_A(z)$ are the Hubble parameter and angular diameter distance at the observed redshift $z$, and $r_d$ is the sound horizon at the drag epoch. The superscripts ``fid'' and ``tem'' refer to these quantities evaluated in the fiducial cosmology, such that $\alpha_{\parallel,\perp}$ are the ratio comparing the BAO position along and perpendicular to the line of sight to that expected in the assumed cosmology; given an oscillatory template $P_w^{\rm tem}$ for the BAO, the scale dependence of the observed signal can then be written as $P_w^{\rm obs}(k_\parallel, k_\perp) \sim P_w^{\rm tem}(k_\parallel/\alpha_\parallel, k_\perp/\alpha_\perp)$. The BAO scaling parameters can be recast into an isotropic $\alphaiso \equiv (\apar/\aperp^2)^{1/3}$, characterizing the absolute geometric mean size of the BAO feature on the sky, and an anisotropic $\alphaap\equiv 
\apar/\aperp$ component characterizing angular distortions due to a mismatch between the fiducial and true cosmologies often known as the Alcock-Paczynski effect~\cite{Alcock:1979mp}.


While standard BAO analyses essentially measure the BAO scale by fitting a rescaled linear template to the measured signal, nonlinear structure formation somewhat complicates the picture since it can generate contributions to the observed power spectrum degenerate with shifts in the BAO. Small changes in the BAO scale can be Taylor expanded as $d P_w^{\rm obs} /d \alpha_{\parallel,\perp} \sim  - (k_{\parallel,\perp}/k)^2 d P_{w}^{\rm tem} / d\ln k$---these derivatives are parametrically large, and shifts in the BAO rather detectable, because the log-derivative is of order $(k r_d) P_w$, with $(k r_d) \sim 10$ on the scales on which the BAO is measured. However, it turns out that structure formation can lead to very similar contributions to the power spectrum: at next-to-leading order, these contributions come from the $P^{(22)}$ contribution to the one-loop power spectrum~\cite{Crocce:2007dt,Eisenstein:2006nj,2009PhRvD..80f3508P,Sherwin:2012_BAO,Chen:2024tfp}
\begin{equation}\label{eq:P22}
    P^{(22)}(\kvec) = 2\int_{\pvec} K_2(\pvec, \kvec-\pvec)^2 P_{\rm lin}(\pvec)P_{\rm lin}(\kvec-\pvec)\,,
\end{equation}
where $K_2$ is the quadratic Fourier-space kernel for \Lya as defined in Ref.~\cite{Ivanov2024}.

The BAO shift is sourced by modes with longer wavelengths than the BAO and involves a coupling of the gravitational shift towards overdensities and the nonlinear clustering of the \Lya field. Schematically, these two types of quadratic nonlinearities take the forms $\Psi(\textbf{x}) \nabla \delta(\textbf{x}) $ and $\delta^2(\textbf{x})$, and their cross correlation contributes to $P^{(22)}$ as $\langle \Psi(\textbf{x}) \delta(\textbf{0}) \rangle \xi'(\textbf{x})$. Here, $\Psi$ is the gravitational displacement and will tend to point towards overdensities
\begin{equation}
    \langle \Psi(\textbf{x}) \delta(\textbf{0}) \rangle = - \frac{\textbf{x}}{3} \langle \delta^2 \rangle_{x}
    \label{eqn:shift_term}
\end{equation}
where $\langle \delta^2 \rangle_R$ is the mean square overdensity in spheres of radius $R$ -- the effect of this term on the BAO peak is thus to shift it proportionally to the mean square overdensity on BAO scales $\sigma^2_d$. Conceptually, we may picture that in real space, the BAO feature contracts isotropically around overdensities and expands around underdensities, but the \Lya signal is suppressed around the former, leading to an expanded BAO radius on net.\footnote{See, e.g., Refs.~\cite{2009PhRvD..80f3508P,Sherwin:2012_BAO, McQuinn:2015tva,Blas:2016sfa} for the equivalent picture in galaxies.} Importantly, this shift only depends on modes with wavelengths longer than $r_d$, since shorter modes will not expand or contract the BAO feature coherently. The final analytic form for the non-linear BAO shift including redshift-space distortions was derived in Ref.~\cite{Chen:2024tfp}, and we derive the equivalent (and somewhat cumbersome) expression for the forest in Appendix~\ref{app:BAO_shift}.  As we will see, the sign of the expected BAO shift due to this nonlinear coupling effect is rather dependent on the values of the quadratic bias parameters which we can measure from the ACCEL$^2$ simulations.

In order to connect the shift term to measured BAO shifts, we use the Fisher formalism from Ref.~\cite{Chen:2024tfp}. Briefly, the shift in $\apar$ and $\aperp$ are given in this formalism by
\begin{equation}
    \Delta \{\apar,\aperp\} = F^{-1}_{\{\apar,\aperp\}a b} t^b_i C^{-1}_{ij} \epsilon_j\, ,
    \label{eqn:fisher_shifts}
 \end{equation}
where $C$ is the covariance matrix, $t^b_i$ is the linear template $dP_i/d\theta_b$ corresponding to fitting parameter $\theta_b$, and $F_{ab}$ is the Fisher matrix. This setup assumes that, in addition to the BAO scales, we also marginalize over a set of other parameters including the linear density and velocity bias, BAO damping parameters, and a smooth broadband parametrized through a cubic spline basis. We compute $C$ assuming a DESI-like effective volume~\cite{Mcquinn2011} and linear theory power spectrum, and $\epsilon$, are the ``wedges'' defined in total separation $k=(k_{\parallel}^2 +\mathbf{k}_{\perp}^2)^{1/2}$ and in $\mu$ bins for the analytic description of the BAO shift, given in Eq.~\eqref{eq:Pshift}, multiplied by the mean-square linear density fluctuation in a sphere of radius $r_d$ and the response of the power spectrum, $\frac{\dd P}{\dd \ln k}$. The fit is performed with $k_{\rm min}=0.02 \hMpcinv$ and $\kmax = 0.5 \hMpcinv$ with a linear $\Delta k=0.005 \hMpcinv$ spacing. Analogously, we forecast the shift in the BAO parameter for the cross-correlation, $\Delta \{\apar,\aperp\}^{\times}$ of the \Lya forest with quasars~\cite{Font-Ribera:2013fha, duMasdesBourboux:2020pck, DESI_lya_2024}. Since the ACCEL$^2$ simulations do not include a quasar (or halo) catalog, we use best-fit values from the eBOSS survey~\cite{Chudaykin:2022nru}\footnote{See also Ref.~\cite{Simon:2022csv}. } and bias relations from~\cite{Abidi:2018eyd, eBOSS:2017ozs} for an approximate estimate.

\section{Results} \label{sec:results}
In this section, we measure the bias parameters of the one-loop \Lya forest power spectrum in the EFT framework described in Sec.~\ref{sec:LyaEFT}. We fit the theory to measurements of the three-dimensional power spectrum obtained from ACCEL$^2$ simulations, described in Sec.~\ref{sec:simulations}. In Sec.~\ref{sec:results_fits} we present the results from the Monte Carlo exploration of the (bias) parameter space. In Sec.~\ref{sec:results_BAO_shift} we use these results to estimate the non-linear shift in the BAO scaling parameters along and transverse to the line-of-sight for the auto-correlation of the \Lya forest. Further, we present results for the \Lya\ -- quasar cross-correlation using simulation-based and analytic scaling relations for the higher-order biases.

\subsection{EFT parameters from ACCEL$^2$ simulations} \label{sec:results_fits}

\begin{table*}
  \centering
  \renewcommand{\arraystretch}{1.15}
  \begin{tabular}{lcccccc}
  \hline
  \hline
  $b_{\mathcal{O}}$ & \multicolumn{2}{c}{$z=2.0$} &\multicolumn{2}{c}{$z=2.6$} & \multicolumn{2}{c}{$z=3.0$} \\
  &&+stoch. terms&&+stoch. terms&&+stoch. terms
  \\
  \hline
  $b_1$ 
  & $-0.0805 \pm  0.0033$& $-0.0739 \pm  0.0051$
  & $-0.1521 \pm  0.0065$& $-0.1418 \pm  0.0099$
  & $-0.2108^{+0.0098}_{-0.0087}$& $-0.2052 \pm  0.0140$
  \\
  
  $b_{\eta}$ 
  & $\phantom{-}0.1286 \pm  0.0115$& $ \phantom{-}0.1245^{+0.0129}_{-0.0140}$
  & $\phantom{-}0.2298 \pm  0.0211$& $\phantom{-}0.2252 \pm  0.0238$
  & $\phantom{-}0.2992 \pm  0.0294$& $\phantom{-}0.2975 \pm  0.0352$
  \\
  
  $b_2$ 
  & $-0.0079^{+0.0658}_{-0.0392}$& $\phantom{-}0.0504^{+0.0970}_{-0.0734}$
  & $-0.0320^{+0.1533}_{-0.0891}$& $\phantom{-}0.2159^{+0.1819}_{-0.1369}$
  & $-0.0722^{+0.2778}_{-0.1699}$& $\phantom{-}0.2509^{+0.2834}_{-0.2129}$
  \\
  
  $b_{\mathcal{G}_2}$ 
  &$\phantom{-}0.0530^{+0.0711}_{-0.0804}$ & $-0.1684^{+0.1039}_{-0.0765}$
  &$-0.0853^{+0.1267}_{-0.1389}$& $-0.3640^{+0.1630}_{-0.1483}$
  &$-0.2422^{+0.1711}_{-0.1852}$ & $-0.3434 \pm  0.2085$

  \\
  
  $b_{\eta^2}$ 
  &$-0.3467^{+0.0817}_{-0.1418}$ & $-1.1780^{+0.2389}_{-0.2645}$
  &$-0.4341^{+0.1837}_{-0.3151}$ & $-1.5934^{+0.4803}_{-0.5500}$
  &$-0.4504^{+0.2634}_{-0.4639}$ & $-1.0861^{+0.4418}_{-0.7115}$

  \\
  
  $b_{\delta \eta}$ 
  &$-0.3079 \pm  0.0863$ & $-0.4730^{+0.1421}_{-0.1702}$
  &$-0.3227^{+0.1971}_{-0.1738}$ & $-0.3000 \pm  0.2597$
  & $-0.3991^{+0.4304}_{-0.2172}$ &  $-0.2032^{+0.4140}_{-0.2864}$
  \\
  
  $b_{(KK)_\parallel}$ 
  &$\phantom{-}0.5068^{+0.1207}_{-0.1485}$ & $\phantom{-}0.7508^{+0.1979}_{-0.1782}$
  &$\phantom{-}0.3745^{+0.3211}_{-0.3796}$ & $\phantom{-}0.9295 \pm  0.4125$
  &$\phantom{-}0.3852^{+0.5434}_{-0.8292}$ & $\phantom{-}0.5254^{+0.5205}_{-0.7398}$
  \\
  
  $b_{\Pi^{[2]}_\parallel}$ 
  & $-0.1468^{+0.0831}_{-0.0611}$& $ \phantom{-}0.1464 \pm  0.0988$
  & $-0.0316^{+0.1544}_{-0.1213}$& $\phantom{-}0.4937^{+0.2136}_{-0.1884}$
  & $\phantom{-}0.0130^{+0.2527}_{-0.1787}$& $\phantom{-}0.3782^{+0.3102}_{-0.2382}$
  \\ \hline
  
  $b_{\Pi^{[3]}_\parallel}$ 
  &$ \phantom{-}0.7923 \pm 0.0193 $ & $  \phantom{-}0.5863 \pm 0.0393 $
  &$  \phantom{-}1.0014 \pm 0.0517 $ & $  \phantom{-}1.0935 \pm 0.1014 $
  &$  \phantom{-}1.1355 \pm 0.0871 $ & $  \phantom{-}1.0831 \pm 0.1669 $
  \\
  
  $b_{\delta\Pi^{[2]}_\parallel}$ 
  &$  \phantom{-}1.4481 \pm 0.0466 $ & $  \phantom{-}1.6968 \pm 0.0685 $
  &$ \phantom{-}0.5003 \pm 0.1200 $ & $ \phantom{-}0.6907 \pm 0.1737 $
  &$ -0.0774 \pm 0.2036 $ & $ \phantom{-}0.4733 \pm 0.2754 $
  \\
  
  $b_{(K\Pi^{[2]})_\parallel}$ 
  &$ -1.5540 \pm 0.0371 $ & $ -1.0424 \pm 0.0943 $
  &$ -1.5008 \pm 0.0948 $ & $ -3.0860 \pm 0.2409 $
  &$ -1.5125 \pm 0.1546 $ & $ -1.8050 \pm 0.3874 $
  \\
  
  $b_{\eta\Pi^{[2]}_\parallel}$ 
  &$ \phantom{-}4.3442 \pm 0.0968 $ & $  \phantom{-}5.1990 \pm 0.1273 $
  &$ \phantom{-}1.8034 \pm 0.2504 $ & $  \phantom{-}3.2778 \pm 0.3233 $
  &$ \phantom{-}0.5513 \pm 0.4273 $& $   \phantom{-}1.3736 \pm 0.5138 $
  \\
  
  $c_0$ 
  &$ -0.0211 \pm 0.0004 $ & $ -0.1557 \pm 0.0072 $
  &$ -0.0163 \pm 0.0006 $ & $ -0.1638 \pm 0.0135 $
  &$ -0.0200 \pm 0.0009 $ & $ -0.1147 \pm 0.0212 $
  \\
  
  $c_2$ 
  &$ \phantom{-}0.0385 \pm 0.0020 $ & $  \phantom{-}0.2554 \pm 0.0086 $
  &$ \phantom{-}0.0456 \pm 0.0037 $ & $ \phantom{-}0.2662 \pm 0.0155 $
  &$ \phantom{-}0.0432 \pm 0.0049 $ &  $ \phantom{-}0.1655 \pm 0.0219 $
  \\
  
  $c_4$ 
  &$ -0.0214 \pm 0.0029 $ & $ -0.1675 \pm 0.0055 $
  &$ -0.0427 \pm 0.0054 $ & $ -0.1920 \pm 0.0098 $
  &$ -0.0399 \pm 0.0075 $ & $ -0.1289 \pm 0.0136 $
  \\

  $P_{\rm shot}$ 
  &- & $\phantom{-}0.1562 \pm 0.0107 $
  &- & $\phantom{-}0.2378 \pm 0.0273 $
  &- & $\phantom{-}0.2109 \pm 0.0506 $
  \\
  
  $a_0$ 
  &- & $ -0.1156 \pm 0.0101 $
  & - & $ -0.1940 \pm 0.0250 $
  &- & $ -0.1795 \pm 0.0452 $
  \\  
  $a_2$ 
  & - & $ -0.0409 \pm 0.0330 $
  & - & $ -0.0026 \pm 0.0816 $
  & - &  $ \phantom{-}0.0672 \pm 0.1249 $
  \\
  $\chi^2$ 
  &$239.1$ & $222.22$
  &$246.6$ & $233.6$
  &$224.6$ & $220.9$
  \\
  $\chi^2_{\nu}$ 
  &$1.27$ & $1.20$
  &$1.31$ & $1.26$
  &$1.19$ & $1.19$
  \\
  \hline
  \end{tabular}
  \caption{Mean best-fit values for the one-loop EFT parameters obtained from the ACCEL$^2$ simulations. The default fit is performed with $\kmax=2\hMpcinv$. The counterterms are divided by $(\hMpcinv)^2$. We analytically marginalize over the parameters shown in the  bottom part of the table and recover their posteriors from the chains \textit{a posteriori}. The resulting $\chi^2$ and reduced $\chi^2_{\nu}$ for the best-fit linear parameters are quoted in the last two rows for 203 data points and 15 (18) degrees of freedom.}
  \label{tab:bestfit_table_lowz}
\end{table*}

\begin{table*}
  \centering
  \renewcommand{\arraystretch}{1.15}
  \begin{tabular}{lccccc}
  \hline
  \hline
  $b_{\mathcal{O}}$ & \multicolumn{2}{c}{$z=3.6$} & \multicolumn{2}{c}{$z=4.0$} & \\
  &&+stoch. terms&&+stoch. terms&
  \\
  \hline
  $b_1$ 
  & $-0.3142^{+0.0154}_{-0.0136}$& $-0.2974^{+0.0253}_{-0.0198}$
  & $-0.3950^{+0.0201}_{-0.0167}$& $-0.3862^{+0.0314}_{-0.0234}$
  \\
  
  $b_{\eta}$ 
  & $\phantom{-}0.3743 \pm  0.0408$ &  $\phantom{-}0.3932 \pm  0.0530$
  & $\phantom{-}0.4082 \pm  0.0473$&  $\phantom{-}0.4112 \pm  0.0598$
  \\
  
  $b_2$ 
  &$\phantom{-}0.0164^{+0.4481}_{-0.3943}$ & $\phantom{-}0.2422^{+0.3329}_{-0.8251}$
  & $\phantom{-}0.2249^{+0.4369}_{-0.7629}$ & $\phantom{-}0.3541^{+0.4389}_{-1.2618}$
  \\
  
  $b_{\mathcal{G}_2}$ 
  &$-0.5030 \pm  0.2781$& $-0.4067^{+0.2473}_{-0.4829}$
  &$-0.5166^{+0.3085}_{-0.4640}$ & $-0.5137^{+0.3333}_{-0.7027}$
  \\
  
  $b_{\eta^2}$ 
  &$-0.6687^{+0.3688}_{-0.6262}$ & $-0.8700^{+1.0507}_{-0.4853}$
  &$-0.5246^{+0.7491}_{-0.5025}$ & $-0.7106^{+1.3443}_{-0.4617}$
  \\
  
  $b_{\delta \eta}$ 
  & $-0.6103^{+0.8186}_{-0.3743}$& $-0.6095^{+0.8900}_{-0.4426}$
  &$-0.7425^{+0.9396}_{-0.5099}$ & $-0.6185^{+1.0382}_{-0.5783}$
  \\
  
  $b_{(KK)_\parallel}$ 
  &$\phantom{-}0.6148^{+1.0023}_{-1.3706}$ & $\phantom{-}0.7192^{+0.9197}_{-1.4181}$
  & $\phantom{-}0.2935^{+1.2036}_{-1.4516}$& $\phantom{-}0.8913^{+1.1315}_{-1.5691}$
  \\
  
  $b_{\Pi^{[2]}_\parallel}$ 
  &$\phantom{-}0.0069^{+0.3535}_{-0.2401}$ & $\phantom{-}0.1193^{+0.4783}_{-0.2767}$
  &$-0.1387^{+0.4191}_{-0.3092}$ &$-0.0062^{+0.5544}_{-0.3290}$
  \\ \hline
  
  $b_{\Pi^{[3]}_\parallel}$ 
  &$ \phantom{-}0.9870 \pm 0.1563 $ & $ \phantom{-}0.8362 \pm 0.2967 $
  &$ \phantom{-}0.7374 \pm 0.2193 $ & $ \phantom{-}0.6420 \pm 0.4255 $
  \\
  
  $b_{\delta\Pi^{[2]}_\parallel}$ 
  &$ -0.2292 \pm 0.3788 $  & $ -0.5208 \pm 0.4756 $
  &$ -0.7743 \pm 0.5155 $ & $ -0.4412 \pm 0.6095 $
  \\
  
  $b_{(K\Pi^{[2]})_\parallel}$ 
  & $ -1.0865 \pm 0.2868 $& $ -1.9565 \pm 0.6609 $
  & $ -0.3358 \pm 0.4053 $&  $ -1.5090 \pm 0.9068 $
  \\
  
  $b_{\eta\Pi^{[2]}_\parallel}$ 
  & $ \phantom{-}0.6979 \pm 0.7938 $& $  1.0923 \pm 0.9012 $
  & $ \phantom{-}0.2458 \pm 1.0806 $& $  1.0748 \pm 1.1555 $
  \\
  
  $c_0$ 
  &$ -0.0315 \pm 0.0013 $ & $ -0.0365 \pm 0.0375 $
  &$ -0.0336 \pm 0.0016 $ &$ -0.0329 \pm 0.0623 $
  \\
  
  $c_2$ 
  &$ \phantom{-}0.0561 \pm 0.0071 $ & $ \phantom{-}0.0509 \pm 0.0350 $
  &$ \phantom{-}0.0471 \pm 0.0088 $ &$ \phantom{-}0.0348 \pm 0.0487 $
  \\
  
  $c_4$ 
  &$ -0.0583 \pm 0.0106 $ & $ -0.0582 \pm 0.0210 $
  &$ -0.0499 \pm 0.0124 $ & $ -0.0473 \pm 0.0263 $
  \\

  $P_{\rm shot}$ 
  & -& $ \phantom{-}0.0250 \pm 0.1004 $
  & -& $ \phantom{-}0.0145 \pm 0.1851 $
  \\
  
  $a_0$ 
  & -& $ -0.0763 \pm 0.0905 $
  & -&  $ -0.0936 \pm 0.1635 $
  \\  
  $a_2$ 
  & -& $ \phantom{-}0.0562 \pm 0.1880 $
  & -& $ \phantom{-}0.0864 \pm 0.2396 $
  \\
  $\chi^2$ 
  &$215.6$ & $213.9 $
  &$219.8 $ &$ 218.3$ 
  \\
  $\chi^2_{\nu}$ 
  &$1.15$ & $1.15$
  &$1.17$ &$ 1.18$ 
  \\
  \hline
  \end{tabular}
  \caption{Same as Tab.~\ref{tab:bestfit_table_lowz} for the two snapshots at $z=3.6$ and $z=4.0$.}
  \label{tab:bestfit_table_highz}
\end{table*}

Tables~\ref{tab:bestfit_table_lowz} and \ref{tab:bestfit_table_highz} summarize the results of fitting the one-loop \Lya forest EFT model to snapshots of the ACCEL$^2$ simulations at redshifts $z=2.0,2.6,3.0$ and $z=3.6,4.0$, respectively. The fits for the key results are performed up to a maximum wavenumber $\kmax = 2 \hMpcinv$. We quote the best-fit parameters together with the uncertainties for the eight sampled parameters. In the bottom part of both tables, we recover the posteriors from the chains for the parameters that we analytically marginalize over. For each redshift, we compare the effect of adding  stochastic terms (denoted by ``$+\mathrm{st.}$'') to the model, described in Sec.~\ref{sec:LyaEFT}. These are \textit{very strongly} detected at the $\gg 5 \sigma$ level at $z=2.0,\,2.6,\,3.0$ and are consistent with zero at $z=3.6,\,4.0$, following baseline expectations that the two-loop contributions become negligibly small at higher redshift. The bottom two rows quote the $\chi^2$ and reduced $\chi^2_{\nu}$ for 203 data points and 15 (18 when including stochastic terms) degrees of freedom. Including the stochastic terms, consistently improves the (reduced) $\chi^2$. Further, going to higher redshifts improves the $\chi^2$ given the smaller degree of non-linearity which will be discussed later in this section. 

We test our analysis for both projection effects: first, prior volume effects which occurs when the data is not constraining enough for the high-dimensional EFT bias parameter space, and, second, prior weight effects when the true value of a parameter is ruled out (or highly disfavored) by the prior. We do not find evidence for both effects here by finding (i) agreement between the mean and the maximum of the posterior (MAP) of each chain; and (ii) we find negligible differences in MAP values when using uninformative, flat or Gaussian priors.

\begin{figure}
    \centering
    \includegraphics[width=\linewidth]{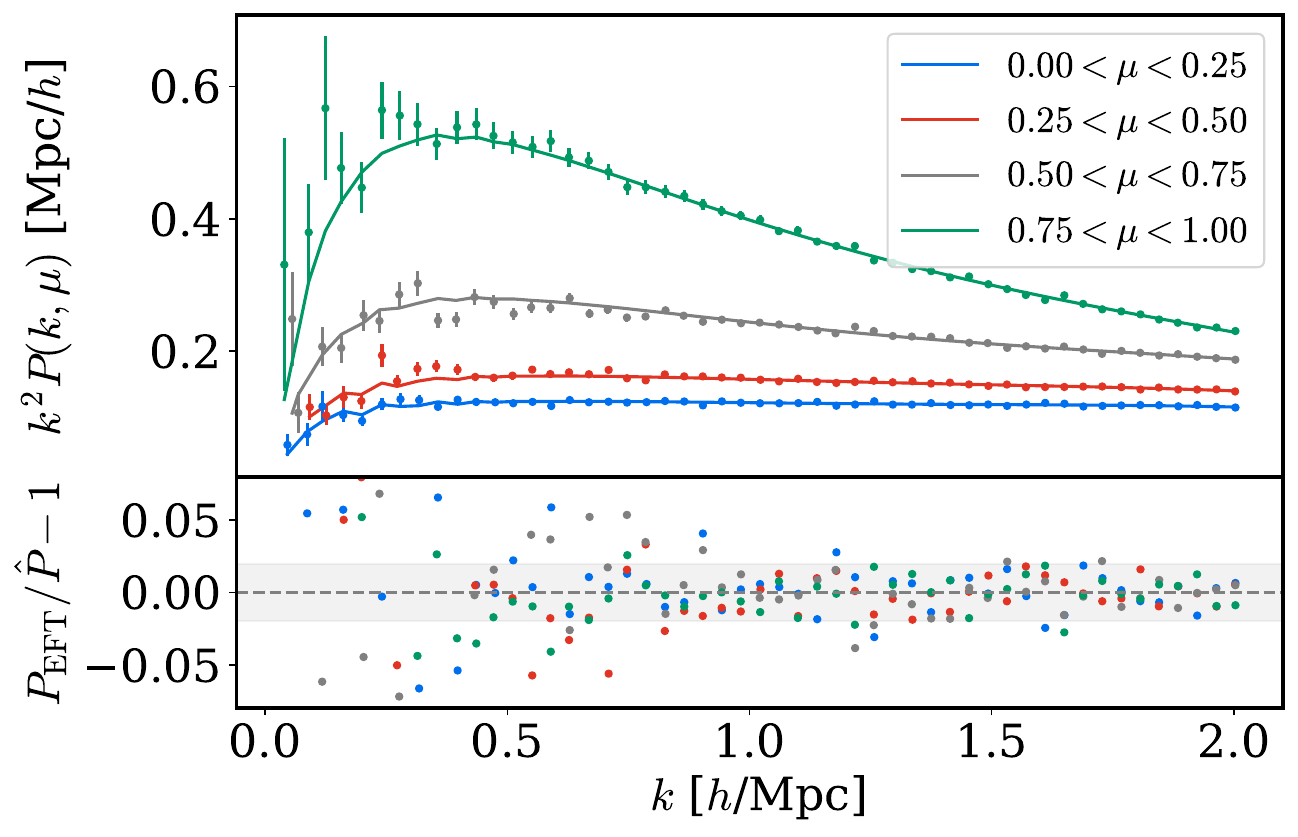}\\
    \includegraphics[width=\linewidth]{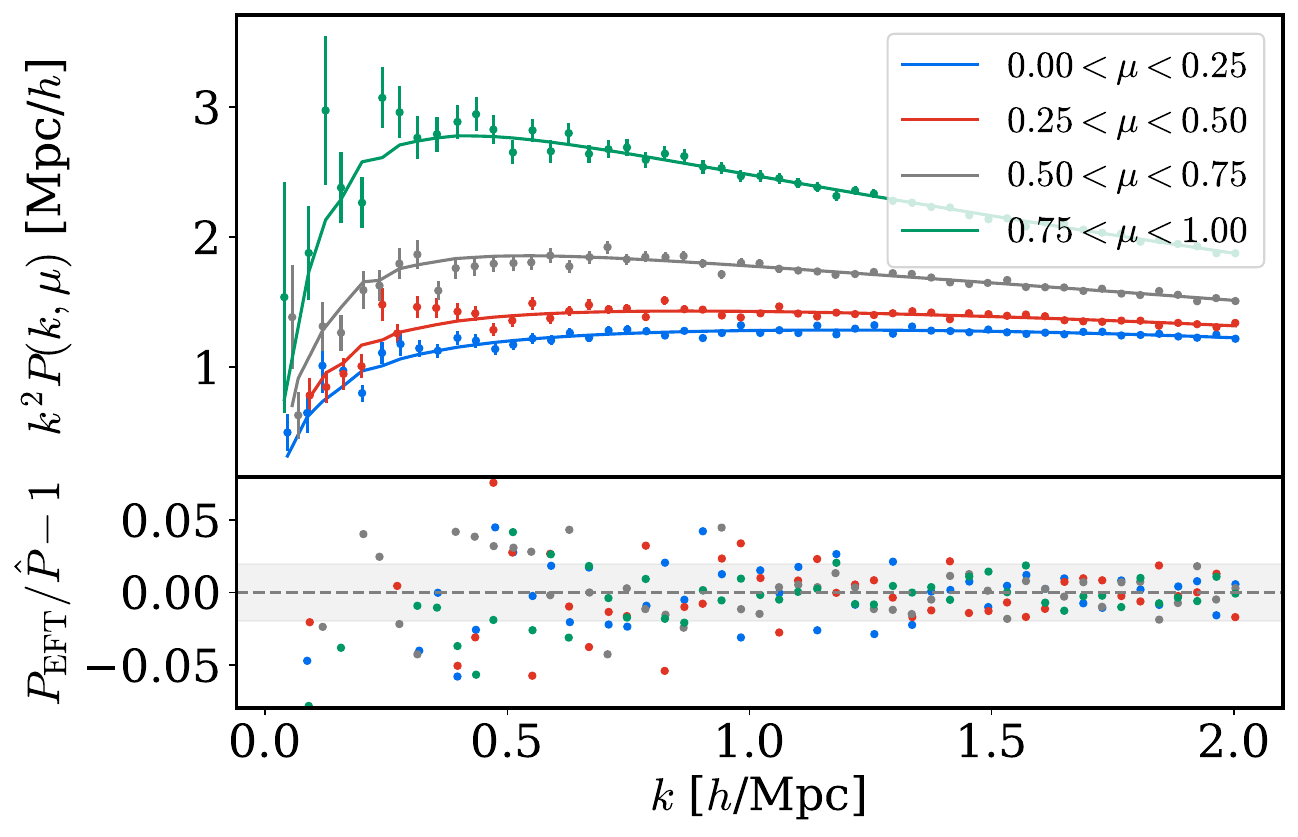}
    \caption{The best-fit EFT model in four angular bins $\mu$ compared to the measured power spectrum from a $z=2.0$ at the top ($z=4.0$ at the bottom) snapshot (top panels), and the residuals between the model and the data (bottom panels). We use $\kmax =2\hMpcinv$ and include stochastic terms. The gray band indicates the 2\% error band to guide the eye.}
    \label{fig:bestfit_pk}
\end{figure}

The resulting best-fit power spectra are shown in Fig.~\ref{fig:bestfit_pk} for the lowest ($z=2.0$) and highest ($z=4.0$) redshift snapshot. Each line represents a ``wedge'' of the power spectrum $P(k,\mu)$ taken from~\cite{Chabanier:2024knr}. The bottom panel shows the residuals reaching sub-$2\%$ accuracy at small scales for $k\simgt 1\hMpcinv $. On large scales ($k\simlt 0.5\hMpcinv$), the residuals approach the $5-10\%$ level. Note that the large-scale fits closest to the line of sight are slightly biased low because we do not introduce a noise floor resulting in constraints predominantly coming from small scales. For comparison, in Ref.~\cite{Chabanier:2024knr} the residuals at large scales reach 20\% which, in our case, are of order 10\%. 

\begin{figure}
    \centering
    \includegraphics[width=\linewidth]{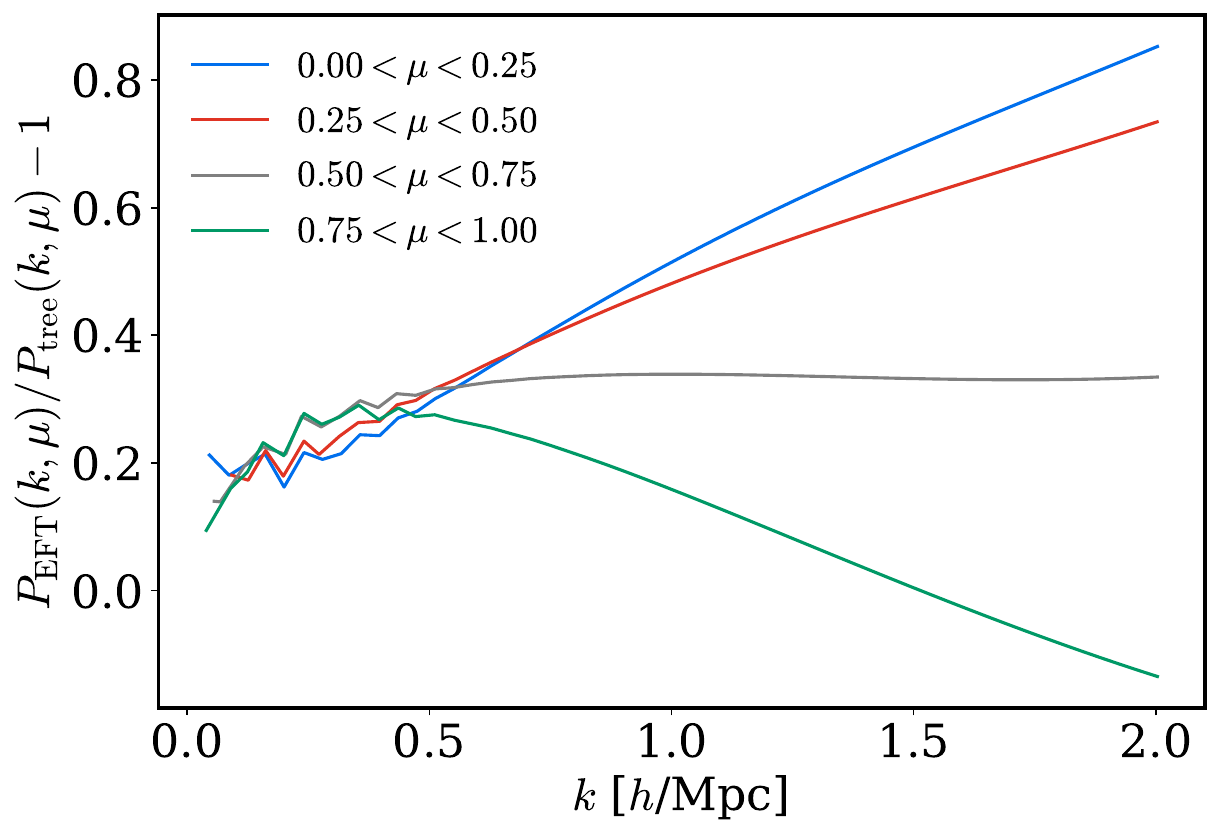}\\
    \includegraphics[width=\linewidth]{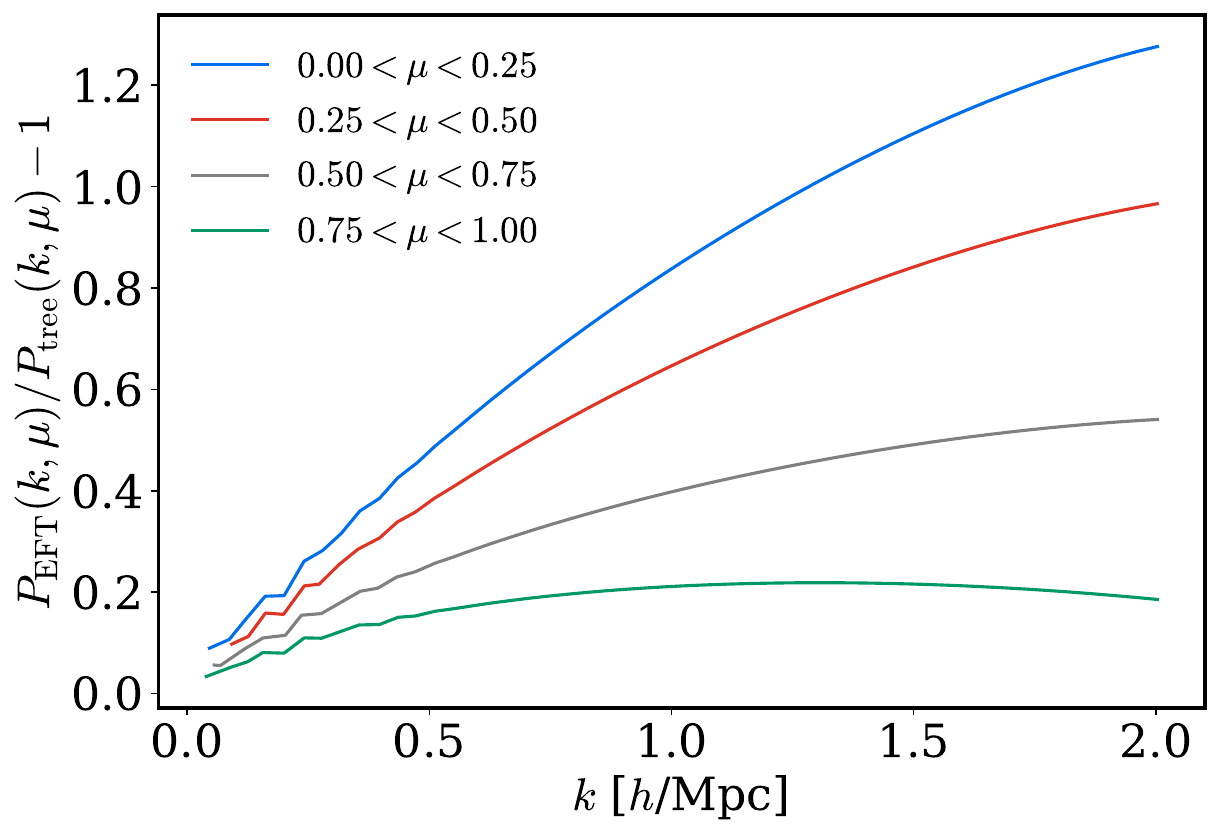}
    \caption{Size of the one-loop corrections obtained by comparing the IR-resummed linear theory power spectrum, $P_{\rm tree}$, to the measured one-loop EFT power spectrum. The top (bottom) plot is a snapshot at $z=2.0$ ($z=4.0$). The corresponding best-fit values are shown in Fig.~\ref{fig:bestfit_pk}. With increasing redshift, the degree of non-linearity decreases which, in turn, increases the perturbative reach of the presented EFT. 
    }    \label{fig:loop_correction}
\end{figure}

\begin{figure*}
\centering
\includegraphics[width=0.49\textwidth]{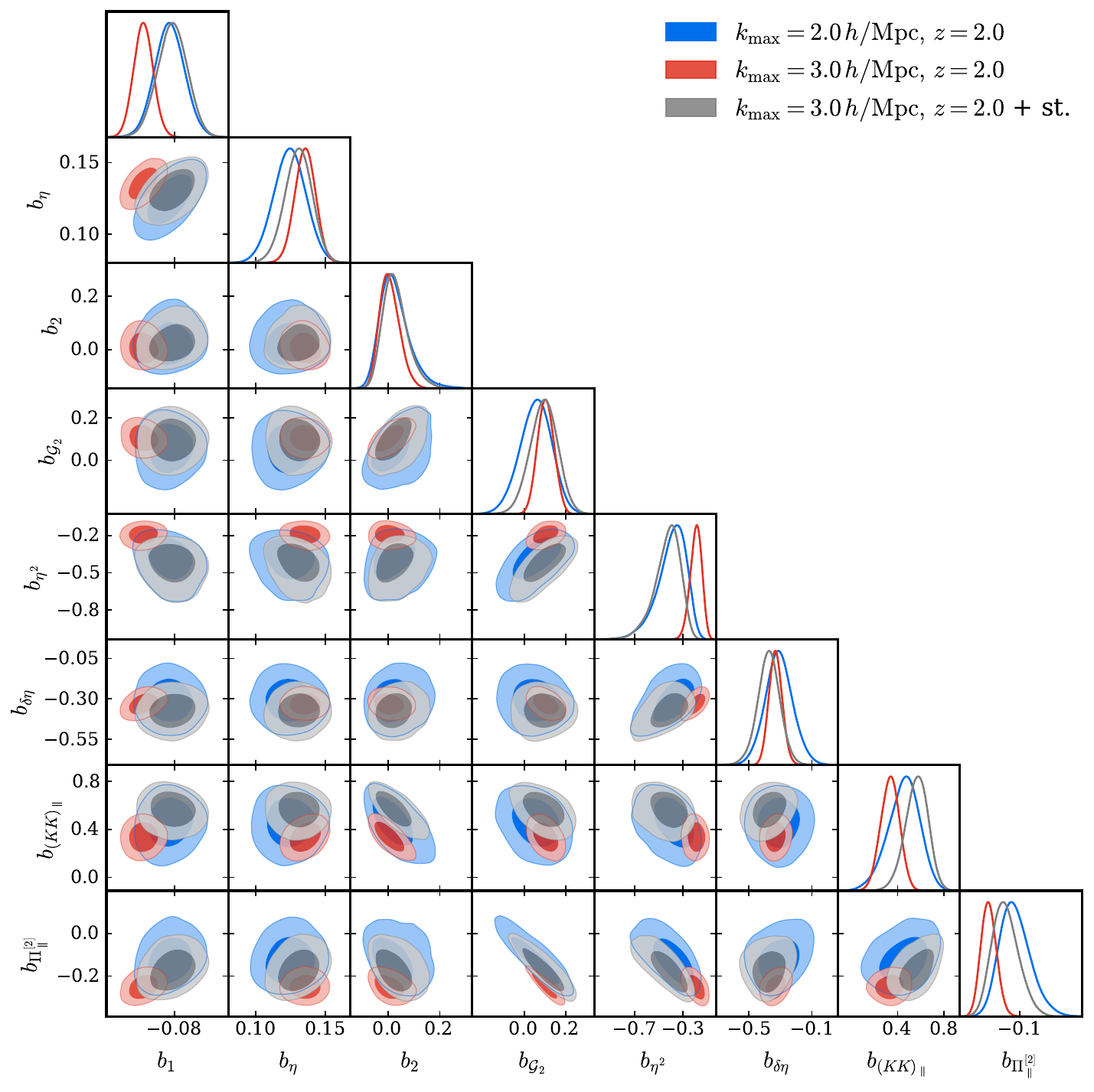}\hfill
\includegraphics[width=0.49\textwidth]{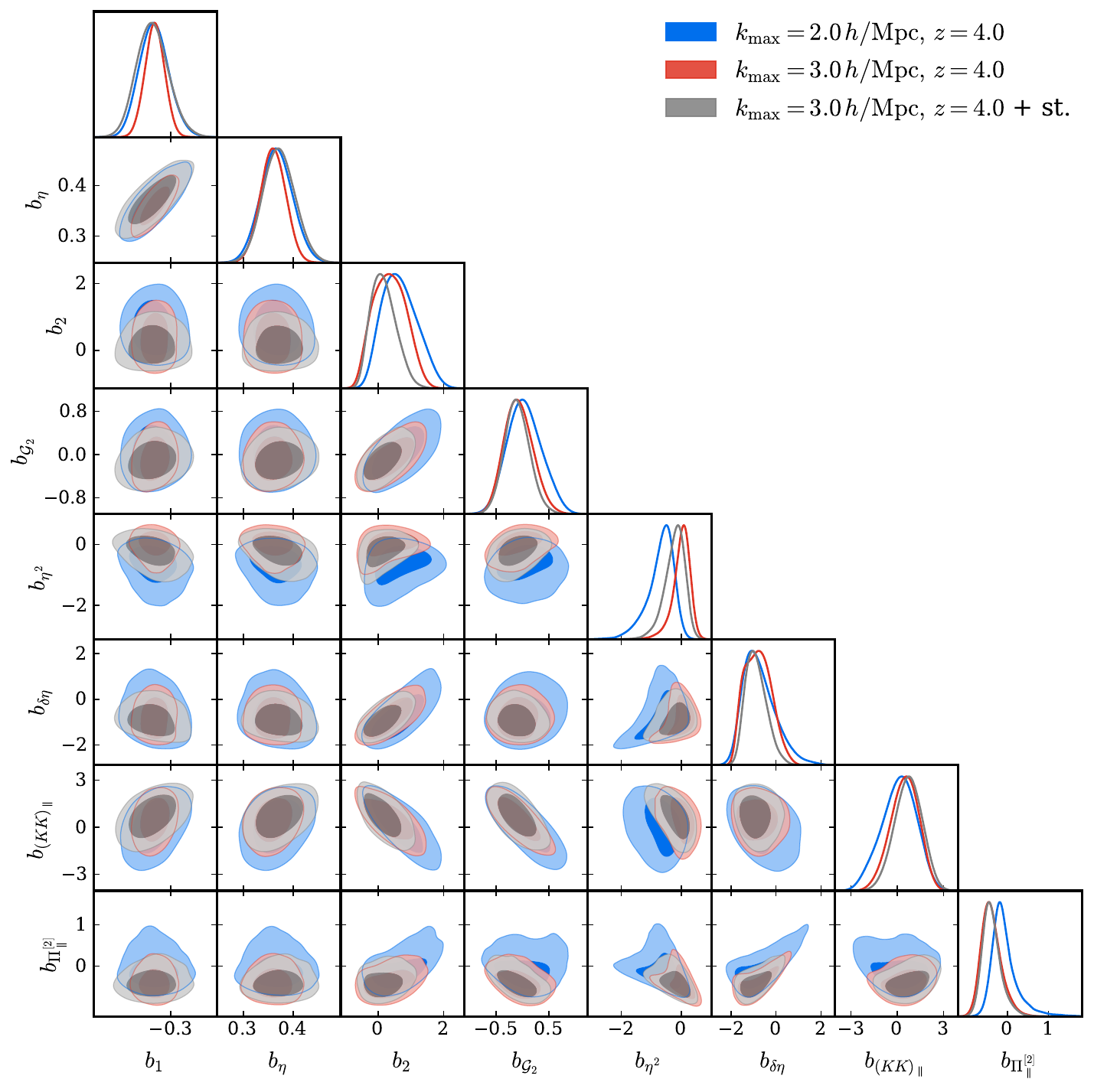}
\caption{Bias parameters and counterterms fit to the ACCEL$^2$ simulations. We show the best-fit 1D and 2D marginalized posteriors for two redshifts $z=2$ (left) and $z=4$ (right). We compare two maximum wavenumbers of $k_{\rm max}=2 \hMpcinv$ and $3$ in the fits. The fits are consistent when including stochastic terms (st.).} \label{fig:triangle}
\end{figure*}

In Fig.~\ref{fig:loop_correction} we show the size of the loop corrections, \emph{i.e.}~the ratio of the EFT power spectrum, $P_{\rm EFT}$, to the IR-resummed tree-level, $P_{\rm tree}$, one. 
The corrections for $z=4.0$ on large scales are smaller than for $z=2.0$, following baseline expectation. 
Further, the unity-crossing of the loop corrections can be viewed as indicative when EFT formally breaks down which, for the high redshift snapshot, occurs at $k\sim 1.5 \hMpcinv$.
Note that 
the unity crossing 
happens first to the 
the one-loop contributions transverse to the line-of-sight ($\mu \leq 0.5$ shown in blue and red).
On the one hand, one
may adopt a conservative 
approach that
the EFT model cannot be 
considered a controlled
approximation for
$k\gtrsim 1 \hMpcinv$.
Whilst in this approach
the model needs to be treated as phenomenological, it still yields highly accurate best-fit power spectra down to small scales shown in Fig.~\ref{fig:bestfit_pk}. 
Note however that the drift plot test
displayed in Fig.~\ref{fig:triangle}
does not show any sign of the 
bias induced by the higher order
corrections. This suggests 
an alternative point of view 
that 
the apparent enhancement 
of the one-loop corrections w.r.t.~tree-level results may not 
actually imply a breakdown 
of EFT, but rather be 
a result of the accidental suppression
of the tree-level contributions
by small values of linear bias parameters.
In this picture, 
the two-loop corrections 
may still be suppressed w.r.t.~the one-loop results 
even at $\kmax\sim 2~\hMpcinv$,
without jeopardizing the 
validity of one-loop EFT 
results at these scales.\footnote{There are many examples
of particle physics processes for which the 
tree-level result is zero, but the one-loop amplitude is not: 
the decay of the Higgs boson 
into two photons, 
the photon-photon
scattering in quantum 
electrodynamics etc. 
The vanishing of the tree-level
results in these cases does not 
imply an inconsistency of the 
one-loop calculation.} 
A definitive resolution
of this issue requires
either a complete two-loop
computation, or a comparison
of one-loop EFT with simulations
at the field level, along the lines of~\cite{Schmittfull:2018yuk,Schmittfull:2020trd,Ivanov:2024hgq,Ivanov:2024xgb,Ivanov:2024dgv}.
We defer these analyses to future work.
As we shall see later, our
BAO shift
estimates will not
depend on a particular 
point of view on the 
role of the one-loop
corrections.

For our fits we explore a range of fitting scales: at $\kmax=0.5 \hMpcinv$, we get broad constraints on the tree-level bias parameters with $b_{\eta^2},\, b_{\td\eta}$,\, with $b_{(KK)_{\parallel}},\, b_{\Pi^{[2]}_{\parallel}}$ recovering their priors; when pushing to very small scales at $\kmax = 5 \hMpcinv$, we get consistent results with our baseline analysis (with significantly smaller error bars), though we emphasize that these scales are beyond the non-linear scale where the EFT model can be trusted. As a consistency check, we also compare our baseline results to the $b_1$ and $\beta$ values computed in Ref.~\cite{Chabanier:2024knr}. Their analysis utilizes a heuristic non-linear fitting function derived from hydrodynamical simulations~\cite{Arinyo-i-Prats:2015vqa}. Our findings show good agreement with their results, at the $\sim 1\sigma$-level across all fits. Specifically, for $b_1$ ($\beta$) we are consistent to $1.2,\,0.8,\,0.3,\,0.8,\,0.7 \sigma$ ($0.7,\,0.4,\,0.03,\,0.7,\,0.8\sigma$) in ascending order of redshift.  

In Fig.~\ref{fig:triangle} we show a drift plot for the evolution of the marginalized 1D and 2D posteriors for the eight bias parameters for $z=2$ shown on the left ($z=4$ on the right) for three different configurations using a maximum wavenumber of two (blue) and three (red) for the fits and additionally including stochastic terms (gray). Following baseline expectations, with increasing redshift and, thus, with decreasing degree of non-linearity, we obtain agreement between the different fits. These results are in good qualitative agreement with the drift plots presented in Ref.~\cite{Ivanov2024}.

\begin{figure}
    \centering
    \includegraphics[width=1\linewidth]{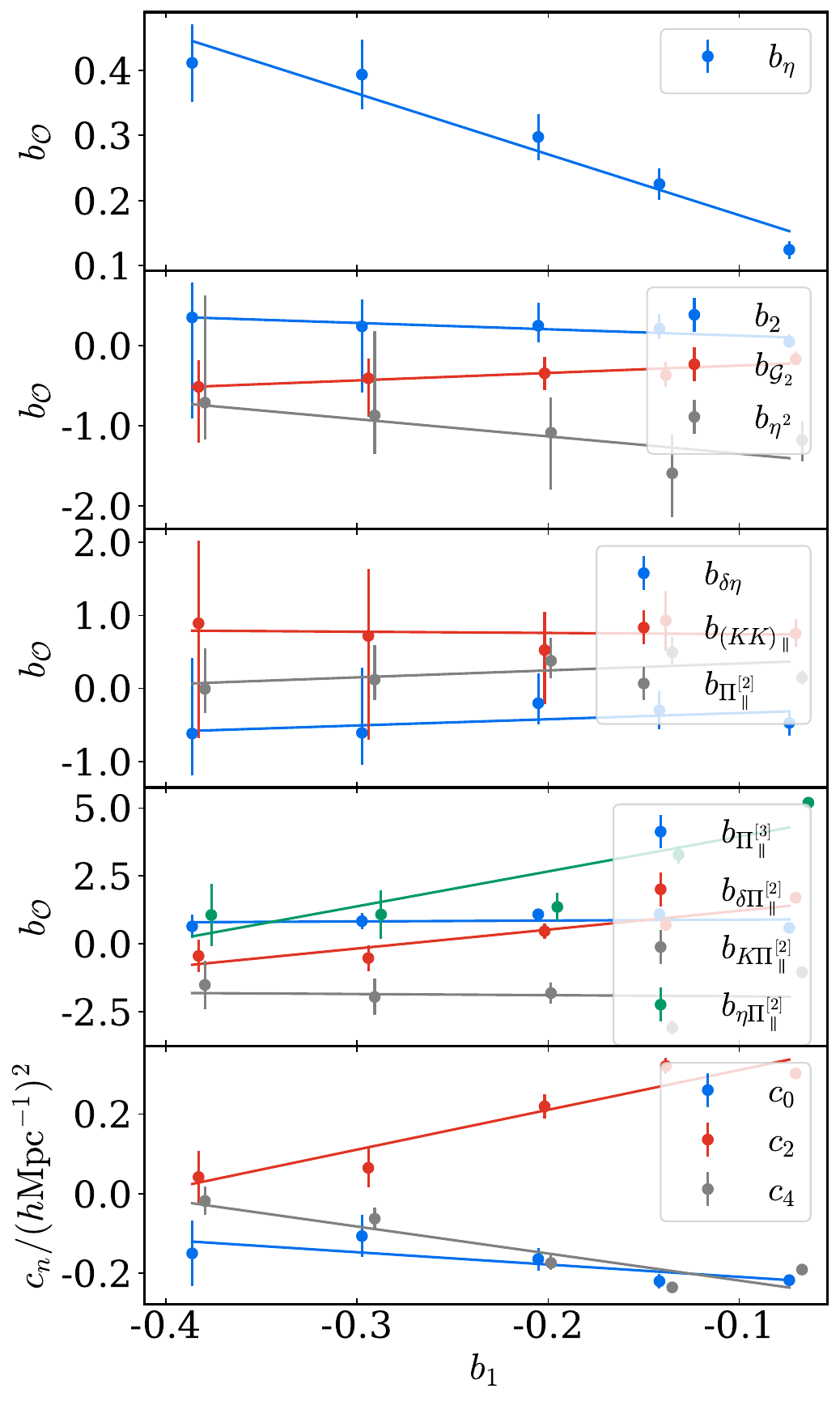}
    \caption{Linear fits (solid lines) to the bias parameter and counterterms. These are obtained from fits to the small ACCEL$^2$ simulation (\texttt{ACCL2\_L160R25}) with $\kmax=2.0\hMpcinv$, tabulated in Tab.~\ref{tab:bestfit_table_lowz} and \ref{tab:bestfit_table_highz}. The linear bias is shown for decreasing redshift (from left to right for $z=4.0, 3.6, 3.0, 2.6, 2.0$). We remind the reader that our convention of $b_{\eta}$ is related to the literature by a negative sign.  }
    \label{fig:eftparam}
\end{figure}

In Fig.~\ref{fig:eftparam} we show the relation of 
the non-linear bias parameters $b_{\mathcal{O}}$ as a function
of 
the linear bias $b_1$. The latter decreases with redshift,
and can be used as a proxy thereof. 
The solid lines are linear fits to the $b_1-b_{\mathcal{O}}$ and $b_1-c_n$ relations as in Ref.~\cite{Ivanov:2024jtl}. 
Error bars for the quadratic and cubic EFT parameters are obtained from the Gaussian covariance and for the analytically marginalized parameters are obtained a posteriori from the chains by sampling from an appropriate Gaussian. We emphasize that the error bars strongly depend on the employed $\kmax$ in the fits. Overall, 
we find good agreement 
with $b_{\mathcal{O}}-b_1$
relations measured 
before in Ref.~\cite{Ivanov:2024jtl}~for the Sherwood 
simulation data~\cite{Bolton17,Givans2022}.

\begin{figure}
    \centering
    \includegraphics[width=1\linewidth]{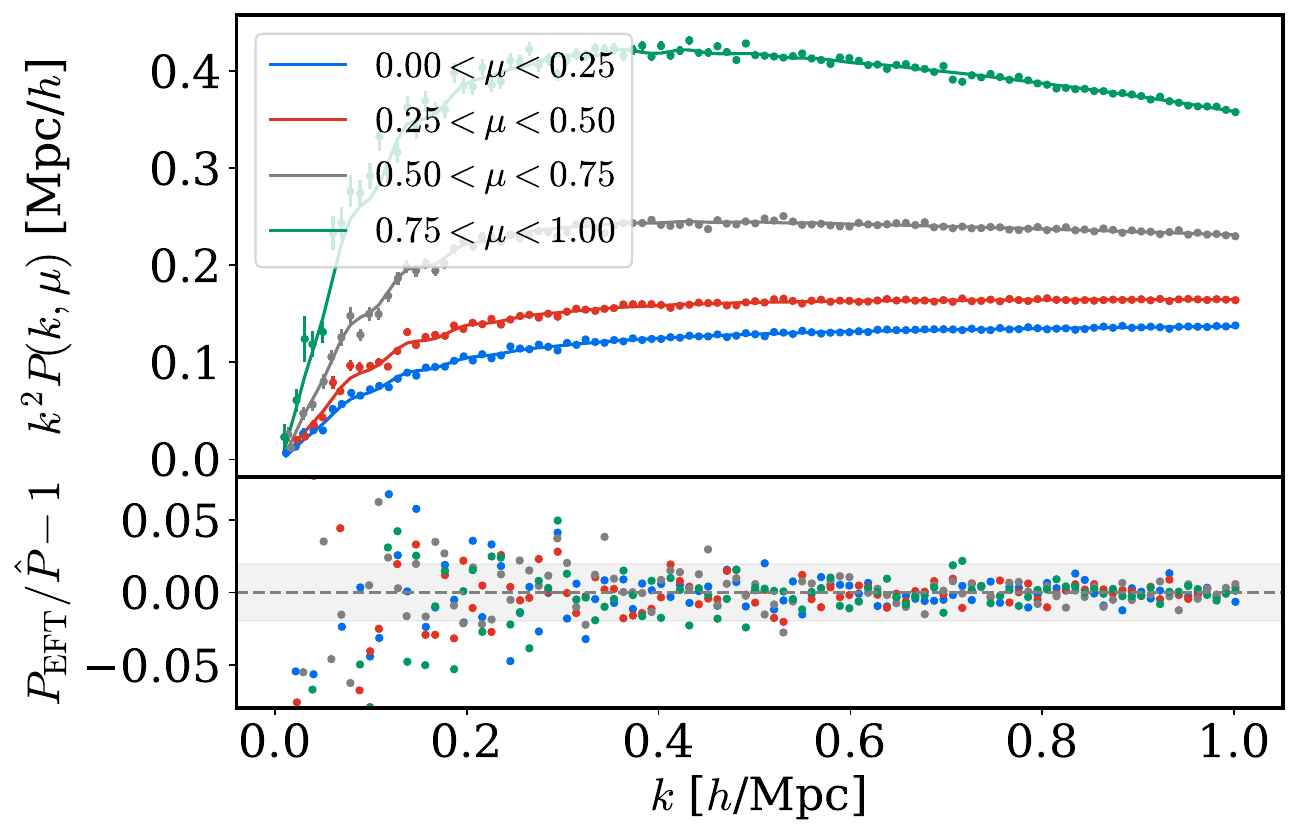}
    \caption{Same as Fig.~\ref{fig:bestfit_pk} for the large box of size $L=640 \hinvMpc$ with an effective resolution of $100\, h^{-1}\mathrm{kpc}$ at redshift $z=2.0$. The gray band indicates the 2\% error band to guide the eye.}
    \label{fig:bestfit_pk_largebox}
\end{figure}

\begin{figure}
    \centering
    \includegraphics[width=1\linewidth]{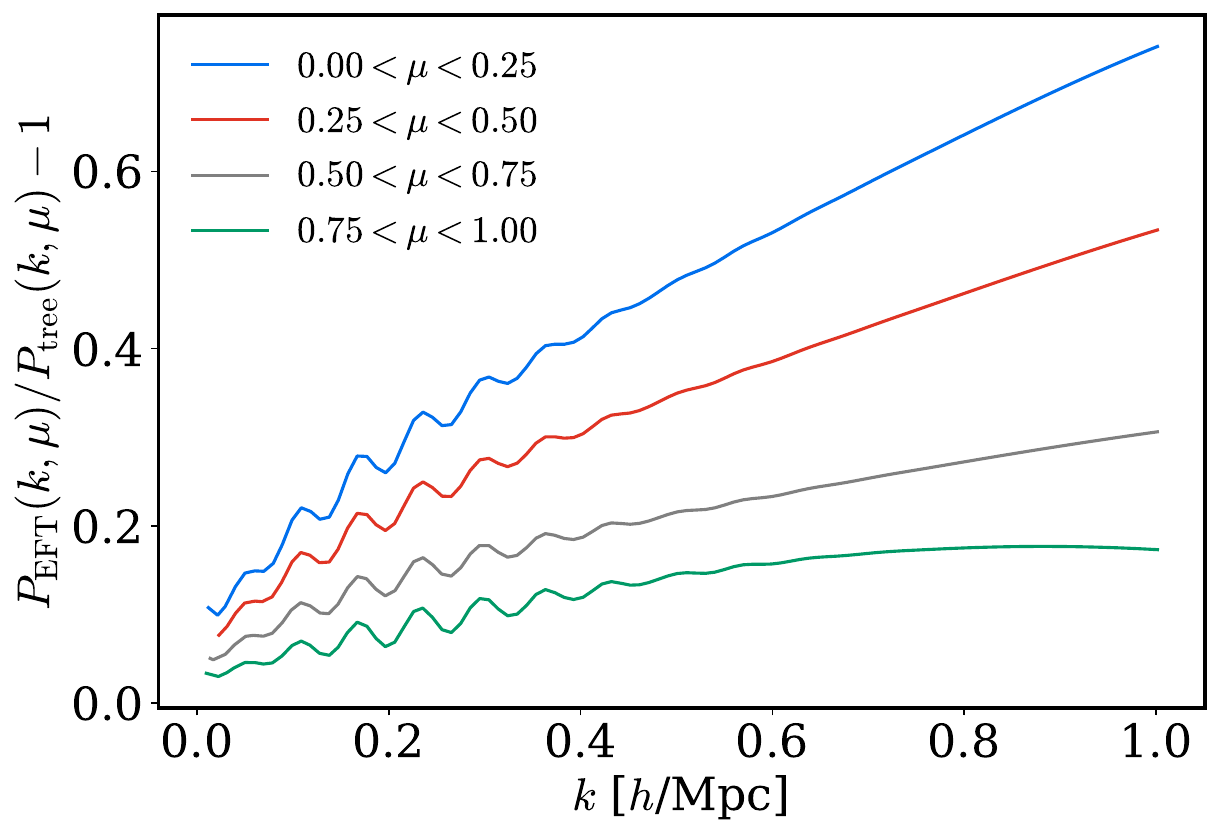}
    \caption{Size of the one-loop corrections obtained by comparing the IR-resummed tree-level power spectrum, $P_{\rm tree}$, to the measured one-loop EFT power spectrum. The fit is done to a snapshot at $z=2.0$. The corresponding best-fit spectra are shown in Fig.~\ref{fig:bestfit_pk_largebox}.}
    \label{fig:loopcorrection_largebox}
\end{figure}

As a consistency check of our analysis, we use the box of size $L=640 \hinvMpc$ with an effective resolution of $100\, h^{-1}\mathrm{kpc}$, denoted by \texttt{ACCL2\_L640R100}. 
While this box has a somewhat worse resolution than our baseline simulation,
it has the advantage of providing access to more linear modes. We perform the same analysis as for the smaller box \texttt{ACCL2\_L160R25} and show in Fig.~\ref{fig:bestfit_pk_largebox} the resulting best fit power spectrum. We use a $\kmax = 1\hMpcinv$ to give more weight to quasi linear, large-scale modes. 
Following baseline expectations, we find consistent, yet tighter, constraints on the EFT bias parameters. Our findings are approximately in agreement with~\cite{Chabanier:2024knr} when comparing the results of the relative differences between both boxes at the $\sim 2\sigma$-level for $b_1$ and at the $ \sim 1.3\sigma$-level for $b_\eta$. In Fig.~\ref{fig:loopcorrection_largebox} we show the corresponding loop corrections. Overall, the picture is similar to our baseline case shown in Fig.~\ref{fig:loop_correction}. 
One may notice slight differences in the relative size of  
the loop corrections seen when comparing 
Fig.~\ref{fig:loopcorrection_largebox} 
and Fig.~\ref{fig:loop_correction}. 
These differences are more pronounced 
for modes with large line-of-sight projections. 
Technically, the difference stems from different 
EFT parameters of the forest in these
two boxes. This difference is consistent 
with variations caused by resolution.
In particular, 
\cite{Chabanier:2024knr} showed that in linear theory resolution
mostly affects the velocity gradient bias. 
Going beyond linear theory, we should 
expect variations at the level 
of the non-linear line-of-sight dependent EFT 
operators. This explains the behavior
of the loop 
corrections observed in Fig.~\ref{fig:loopcorrection_largebox}.
It will be interesting to carry out our analysis
for the large-box simulations 
where the resolution effects are corrected
with splicing~\cite{Chabanier:2024knr}.
We leave this for future work.

\subsection{Non-linear BAO shift} \label{sec:results_BAO_shift}
In this section, we compute the BAO shift parameter in radial ($\apar$) and transverse ($\aperp$) direction from the fits of the one-loop EFT power spectrum to the ACCEL$^2$ simulations, as described in Sec.~\ref{sec:BAO_shift}. For each set of samples in the MCMC chain we compute the BAO shift to obtain an error estimate. Our baseline results are for \Lya parameters constrained using $\kmax =2 \hMpcinv$ at redshifts $z=2.0$ and $z=2.6$ (the two redshifts closest to current \Lya BAO measurements at $z=2.33$~\cite{DESI_lya_2024})  and we find shifts in percent  at $z=2.0$
\begin{align}\label{eq:BAOshift_auto_z20}
    \Delta \apar &= -0.20 \pm 0.09\, \%\,, & \Delta \aperp &= -0.31 \pm 0.11\, \%\,,
\end{align}
and at $z=2.6$
\begin{align} \label{eq:BAOshift_auto_z26}
    \Delta \apar &= -0.16 \pm 0.05\, \%\,, & \Delta \aperp &= -0.20 \pm 0.07\, \%\,.
\end{align}
This translates to an isotropic ($\alphaiso$) and anisotropic ($\alphaap$) shift in the BAO dilation parameter at $z=2.0$ of
\begin{align}\label{eq:BAOshift_auto_z20_aiso}
    \Delta \alphaiso &= -0.28 \pm 0.09\, \%\,, & \Delta \alphaap &= 0.11 \pm 0.07\, \%\,,
\end{align}
and at $z=2.6$ of
\begin{align}\label{eq:BAOshift_auto_z26_aiso}
    \Delta \alphaiso &= -0.18 \pm 0.06\, \%\,, & \Delta \alphaap &=  0.04 \pm 0.06\, \%\,.
\end{align} 

We obtain consistent, yet tighter, shifts in the BAO peak when using the large simulation box. For the radial and transverse scaling parameters, we find a shift at $z=2.0$ of $\Delta \apar = -0.08 \pm 0.05\, \%$, $\Delta \aperp = -0.05 \pm 0.05\, \%$ and at $z=2.6$ of $\Delta \apar = -0.11 \pm 0.05\, \%$, $\Delta \aperp = -0.04 \pm 0.04\, \%$. These shifts correspond to isotropic and anisotropic scaling parameters at the same redshifts: 
$\Delta \alphaiso = -0.03 \pm 0.08\, \%$ and 
$\Delta \alphaap  = -0.06 \pm 0.03\, \%$ at $z=2.0$ and 
$\Delta \alphaiso = -0.07 \pm 0.03\, \%$ and 
$\Delta \alphaap  = -0.07 \pm 0.06\, \%$ at $z=2.6$, respectively. 

\begin{figure}
\centering
\includegraphics[width=0.49\textwidth]{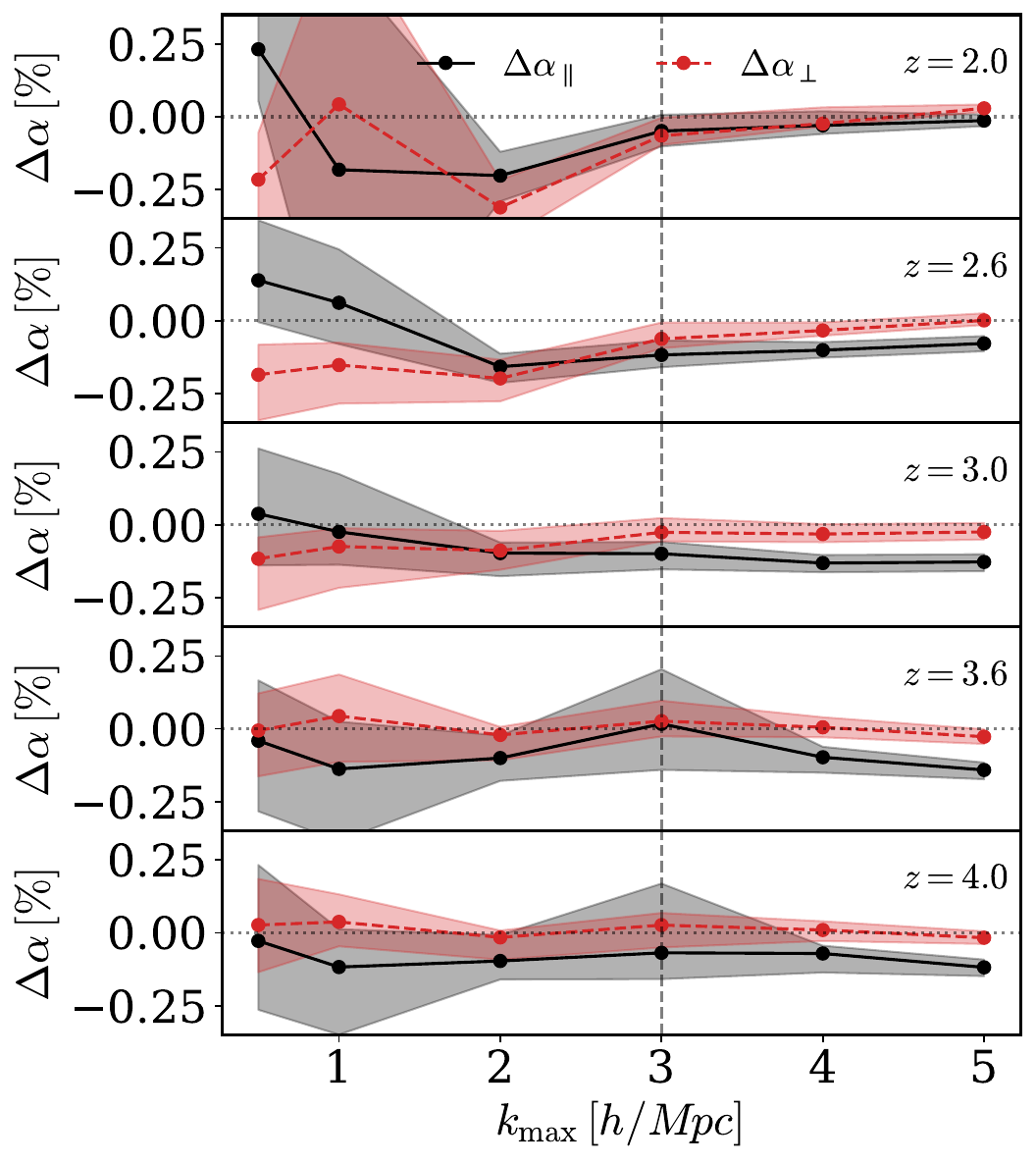}
   \caption{Percentage of shift parameter sampled from the MCMC chains obtained from fitting the one-loop power spectrum to the  ACCEL$^2$ simulations. 
   The bands show the mean and standard deviation in the BAO parameters (relative to an expected value of unity). Redshift is evolving from top to bottom with each simulation snapshot at $z=2.0, 2.6, 3.0, 3.6, 4.0$ and the fitted $\kmax$ ranges from $0.5-5 \hMpcinv$. The vertical dashed line indicates the maximum wavenumber up to which we recommend the use of the BAO shift.} \label{fig:alpha_kmax}
\end{figure}

In Fig.~\ref{fig:alpha_kmax} we show the evolution of the shift parameter as a function of the chosen $\kmax$ in the fits for each snapshot for the radial (transverse) component as a solid black (dashed red) line. The error bars are obtained from computing the shift parameter for each step in the MCMC chain. Following expectations, the uncertainty decreases with increasing $\kmax$. It is interesting to note that the radial shift parameter shows a positive BAO shift but switches towards a negative one at $k\sim 1\hMpcinv$ which is acceptable given the size of the error bars. The transverse shift parameter remains negative or consistent with zero. Whilst we present the evolution of the shift in the BAO peak as a function of maximum wavenumber, we stress that these should only be used up to $\kmax\sim 2-3 \hMpcinv$ for a theoretical and systematic error budget.

In this section, we derive the BAO shift for \Lya forest -- quasar cross-correlation since constraints on the BAO parameters are mostly driven by the cross-correlation (see, Fig.~7 in~\cite{DESI_lya_2024}).\footnote{Cross-correlations of the \Lya forest with quasar positions break the degeneracy of the growth rate $f$ with the (unknown) velocity gradient bias~\cite{Seljak:2012tp,Font-Ribera:2013fha}, stemming from the non-linear mapping of the optical depth to observed flux.} We follow the same procedure as in the \Lya auto-correlation analysis and replace in Eq.~\eqref{eq:P22} one occurrence of the second order redshift space kernel for quasars, i.e. the standard perturbation theory kernel $Z_2$~\cite{Bernardeau:2001CPT,Ivanov:2019pdj}. The resulting analytic form for the shift is given in Appendix \ref{app:BAO_shift}. Since the ACCEL$^2$ simulations do not include a quasar catalog, we follow two approaches to obtain an estimate for the BAO shift: First, using the best-fit values from eBOSS~\cite{Chudaykin:2022nru} and, second, from higher-order quasar bias relations~\cite{eBOSS:2017ozs, Abidi:2018eyd}.

The best-fit values from eBOSS are computed at an effective redshift $z_{\rm eff}=1.48$:
\begin{subequations}
\begin{align} 
    b_1^{q} &= \phantom{-}2.01^{+0.17}_{-0.20} \,, & b_1^{q} &= \phantom{-}1.96^{+0.16}_{-0.20}\,, \label{eq:eboss_b_qso1}\\  
    b_2^{q} &= -0.76^{+0.88}_{-0.93} \,, & b_2^{q} &= -0.28^{+0.91}_{-0.92}\,, \label{eq:eboss_b_qso2}\\ 
    b_{\mathcal{G}_2}^{q} &= -0.31^{+0.37}_{-0.42} \,, & b_{\mathcal{G}_2}^{q} &= -0.18^{+0.42}_{-0.46}\,,\label{eq:eboss_b_qso3}
\end{align}
\end{subequations}
where the left (right) column gives the chosen values for the northern (southern) galactic cap. The resulting BAO shifts at $z=2.0$ are 
\begin{subequations}
\begin{align}
\Delta \apar^{\times} &= -0.10 \, \%\,, & \Delta \aperp^{\times} &= -0.11 \, \%\,, &(\text{NGC})\,, \\
\Delta \apar^{\times} &= -0.15\, \% \,, & \Delta \aperp^{\times} &= -0.15 \, \%\,, &(\text{SGC})\,.
\end{align}
\end{subequations}
We note that the BAO shift is less pronounced in the cross-correlation but caution the reader that the redshift of the quasar biases is not consistent with the simulation and should only be taken indicatively. 

\begin{figure}
    \centering
    \includegraphics[width=1\linewidth]{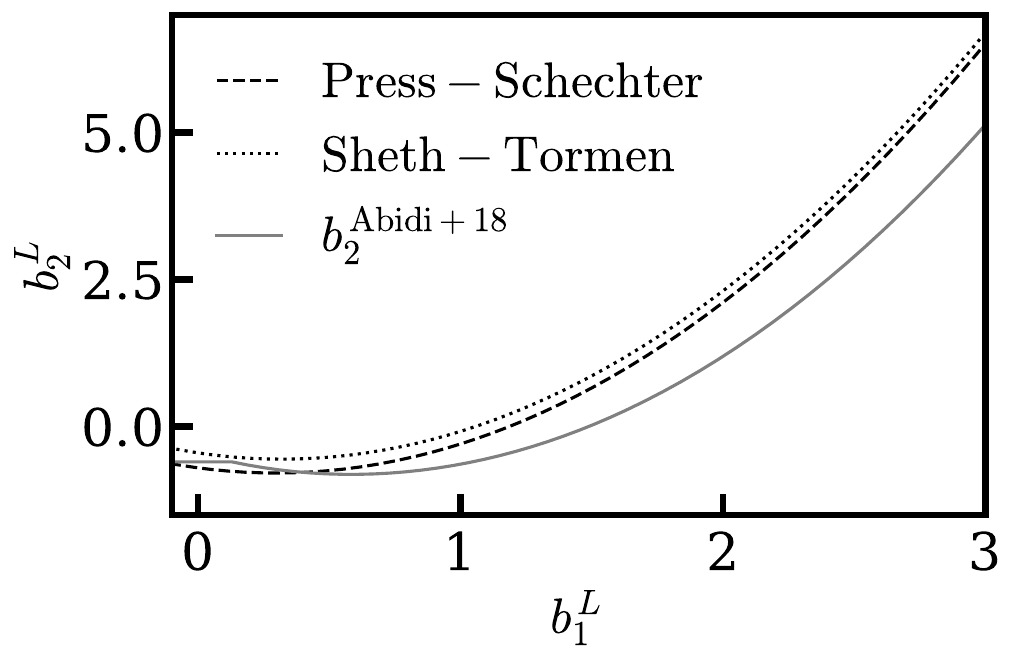}
    \caption{Bias relations assumed for the \Lya--quasar cross-correlation. The two edge cases recovering for $a=1$ and $b=0$ the Press-Schechter (black dashed line) and for $a = 0.707$ and $p = 0.3$ the Sheth-Tormen mass functions (dotted black line) given in Eqs.~\eqref{eq:b1_relation} and \eqref{eq:b2_relation}~\cite{White:2014gfa}. The gray, solid line is the bias relation from~\cite{Abidi:2018eyd} as a function of the quasar bias from~\cite{eBOSS:2017ozs}. Eulerian ($E$) biases are related to Lagrangian ($L$) biases through $b_{1,E} = b_{1,L}+1$. 
    }
    \label{fig:bias_relation}
\end{figure}

To explore the BAO shift of the cross-correlation as a function of redshift, we employ two bias relations: First, for illustrative purposes, we use a simplistic relation for the Lagrangian biases $b_1^L$ and $b_2^L$~\cite{White:2014gfa}
\begin{align}
\label{eq:b1_relation}
b_1^L(\nu) &= \frac{1}{\delta_c} \left( a \nu^2 - 1 + \frac{2p}{1 + (a \nu^2)^p} \right)\,,    \\
\label{eq:b2_relation}
b_2^L(\nu) &= \frac{1}{\delta_c^2} \left( a^2 \nu^4 - 3 a \nu^2 + \frac{2p (2a \nu^2 + 2p - 1)}{1 + (a \nu^2)^p} \right)\,,
\end{align}
where $\td$ is, assuming a peak-background split, the background density with the critical density for collapse given by $\td_c = 1.686$, $\nu$ is the peak height of the density perturbations defined as $\nu = \td_c/\sigma(M)$ where the denominator describes the root mean square of the fluctuations of the smoothed density field over some scale $M$, $a$ is a normalization constant and $p$ a fitting parameter tuned to simulations. Inserting the parameters $a=1.0$ and $p=0.0$ yields the Press-Schechter mass function~\cite{1974ApJ...187..425P}, while $a = 0.707$ and $p = 0.3$ gives the Sheth-Tormen mass function~\cite{1999MNRAS.308..119S}, illustrated in Fig.~\ref{fig:bias_relation}. We set the tidal shear bias $b^L_{s}=0$ and translate the Lagrangian bias parameters above to the Eulerian basis used in the rest of this work via $b_1 = 1 + b_1^L$, $b_2 = b_2^L + \frac43 b_s^L$ and $b_{\mathcal{G}_2}=b_s^L-\frac27 b_1^L$. In order to account for possible nonzero Lagrangian tidal bias we also consider the halo bias relations measured in ~\cite{Abidi:2018eyd}, digitizing the bias values from their Fig.~8. To set the linear bias of quasars we use the measured eBOSS quasar bias relation $b_1(z) = 0.278 \left[(1 + z)^2 - 6.565 \right]+ 2.393$~\cite{eBOSS:2017ozs}. 

\begin{figure}[t!]
\centering
\includegraphics[width=0.49\textwidth]{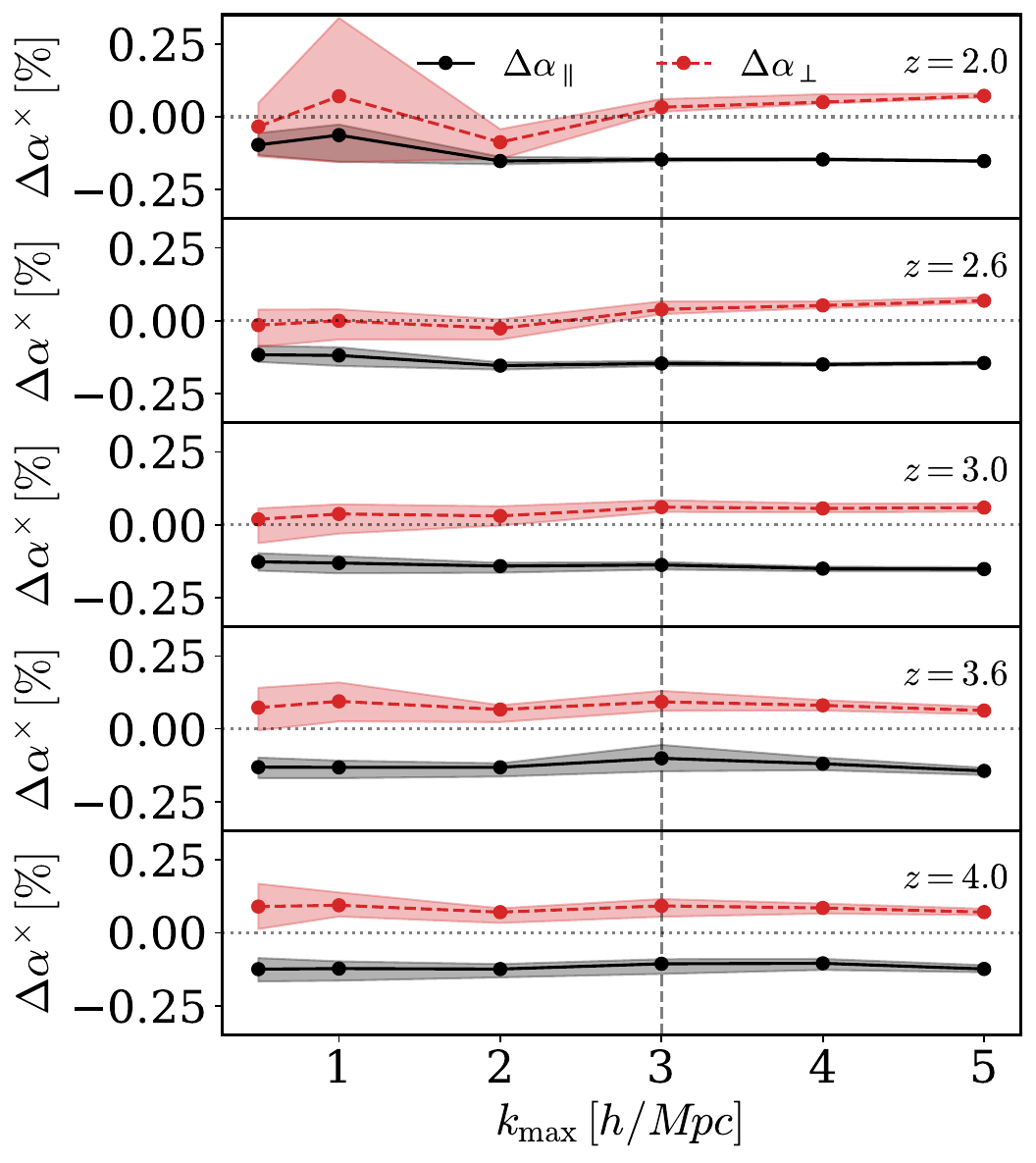}
   \caption{Percentage of BAO shift parameter for the \Lya--quasar cross-correlation, analogous to  Fig.~\ref{fig:alpha_kmax}, using the bias relations from~\cite{Abidi:2018eyd}. The vertical dashed line at $\kmax=3.0\hMpcinv$ indicates the maximum wavenumber up to which we recommend the use of the estimate for the BAO shift. } \label{fig:alphaX_kmax}
\end{figure}

In Fig.~\ref{fig:alphaX_kmax}, we show the BAO shift as a function of the $\kmax$ used in the fits using this bias relation. For the shift in transverse direction we see a similar pattern as in Fig.~\ref{fig:alpha_kmax} that the errors shrink for the first two $\kmax$ values (bar for the snapshot at $z=2.0$) and give shifts that are consistent with zero at the $1-2\sigma$-level. In radial direction that shift has an almost constant offset of $\sim 0.1\%$. 

In summary, for $\kmax =2 \hMpcinv$ at redshifts $z=2.0$ and $z=2.6$ (the two redshifts closest to current BAO measurements from \Lya--quasar cross-correlations at $z=2.33$~\cite{DESI_lya_2024}), we find shifts in percent at $z=2.0$ (first row using the bias-relation from~\cite{Abidi:2018eyd} and second row Sheth-Tormen in the following) of
\begin{subequations}
\begin{align}
    \Delta \apar^{\times} &= -0.15\pm 0.01\,, & \Delta \aperp^{\times} &= -0.09 \pm 0.01\,,\\
    \Delta \apar^{\times} &= -0.09 \pm 0.02\,, & \Delta \aperp^{\times} &= -0.10 \pm 0.05\,,
\end{align}
\end{subequations}
and at $z=2.6$ of
\begin{subequations}
\begin{align}
    \Delta \apar^{\times} &= -0.15 \pm 0.01\,, & \Delta \aperp^{\times} &= -0.03 \pm 0.03\,,\\
    \Delta \apar^{\times} &= -0.12 \pm 0.01\,, & \Delta \aperp^{\times} &= -0.03 \pm 0.03\,.
\end{align}
\end{subequations}
This translates to a shift in percent of the isotropic and anisotropic BAO dilation parameter at $z=2.0$ of
\begin{subequations}
\begin{align}
    \Delta \alphaiso^{\times} &= -0.11 \pm 0.04\,, & \Delta \alphaap^{\times} &= -0.07 \pm 0.05\,,\\
    \Delta \alphaiso^{\times} &= -0.10 \pm 0.03\,, & \Delta 
 \alphaap^{\times} &= \phantom{-}0.01 \pm 0.06\,,
\end{align}
\end{subequations}
and at $z=2.6$ of
\begin{subequations}
\begin{align}
    \Delta \alphaiso^{\times} &= -0.07 \pm 0.03\,, & \Delta \alphaap^{\times} &=  -0.13\pm 0.03 \,,\\
    \Delta \alphaiso^{\times} &= -0.07 \pm 0.03\,, & \Delta \alphaap^{\times} &=  -0.10 \pm 0.03\,,
\end{align} 
\end{subequations}
which yields very similar results for both bias relations. The BAO shift parameters for the cross-correlation indicate that the contribution of the shift to the theoretical and systematic error budget is small (which needs to be added as an error term to the covariance).

\begin{figure}[t!]
    \centering
    \includegraphics[width=\linewidth]{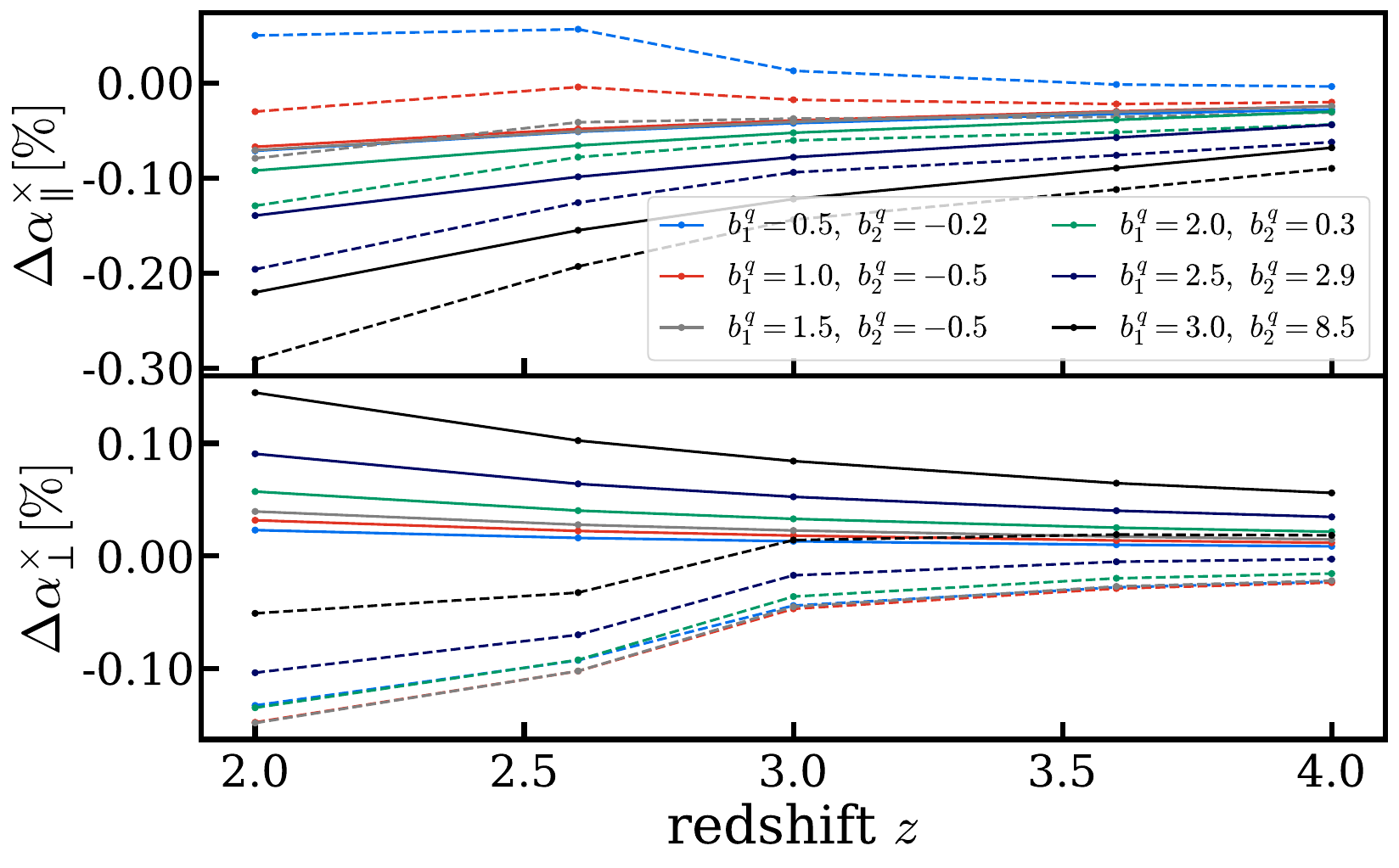}
    \caption{Fisher forecasts for the BAO shift in radial ($\apar$) and transverse ($\aperp$) direction in redshift space for the cross-correlation as a function of redshift $z$. The solid lines only use the best-fit linear biases $b_1,\, b_\eta$ and the dashed lines include the non-linear parameters given in Tabs.~\ref{tab:bestfit_table_lowz}-\ref{tab:bestfit_table_highz}. We vary $b_1^q$ freely and use for $b_2^q$, and $b_{\mathcal{G}_2}^q$ the values obtained from the Sheth-Tormen relation, given in Eq.~\eqref{eq:b2_relation}.}
    \label{fig:BAOshift_cross_intuition}
\end{figure}

To provide the reader with some intuition, we show in Fig.~\ref{fig:BAOshift_cross_intuition} the Fisher forecasts for the BAO shift in radial ($\apar$) and transverse ($\aperp$) direction in redshift space for the cross-correlation as a function of redshift $z$. We compare the case only including linear biases $b_1,\, b_\eta$ (solid lines) and the one including non-linear \Lya bias parameters (dashed lines). The radial shift is slightly increased when including non-linear terms. The transverse shift, however, changes sign and whilst strongly biased quasars (e.g., $b_1^q=3.0, \, b_2^q=8.5$) cause a noticeable shift of 0.15\% that is reduced to 0.05\% when including the non-linear \Lya terms. Schematically, we can think of the BAO shift in the cross correlation as having size
\begin{equation}
    \Delta \alpha \sim \frac{1}{b_1 b_1^q} (b_1 \frak{b}_2^q + b_1^q \frak{b}_2) \sigma^2_d = \left( \frac{\frak{b}_2^q}{b_1^q} + \frac{\frak{b}_2}{b_1} \right) \sigma^2_d
    \label{eqn:cross_shift_schem}
 \end{equation}
where the term in parentheses is the size of the mode-coupling shift term and the denominator is the size of the linear BAO signal, and we have used $\frak{b}_2$ and $\frak{b}^q_2$ to stand in for the amplitude of quadratic nonlinearities in \Lya and quasars. Eq.~\eqref{eqn:cross_shift_schem} indicates that the shift is proportional to the combination of (a) the relative amplitude of the linear and quadratic contributions to clustering; and (b) the amplitude of density fluctuations on BAO scales. While the latter decreases as the growth factor $D(z)^2$ towards higher redshifts, if the corresponding galaxy bias increases the overall BAO shift need not decrease--e.g. for high mass halos we have $b_2^L \approx (b_1^L)^2$, such that the contribution due to quasars in the cross correlation will take the form $b_1^q \sigma^2_d$ for large $b_1^q$, leading to an anomalously large shift amplified by the quasar bias. For the quasar samples in e.g. DESI the linear bias is not sufficiently large to enter this regime--indeed $b_2^L$ tends to be slightly negative, canceling some of the quadratic dynamical nonlinearity--but this effect may be interesting for future surveys involving cross correlations for example with high-redshift Lyman break galaxies (LBGs) \cite{schlegel_megamapper_concept}. 

\section{Summary and Conclusion} \label{sec:conclusion}

Measurements of the baryon acoustic oscillation (BAO) feature are one of the pillars of modern cosmology. High-redshift measurements obtained from the three-dimensional clustering of the \Lya forest at $z \sim 2$ provide measurements of the cosmic expansion in the matter-dominated regime~\cite{DESI_lya_2024}, yielding complimentary information to low-redshift galaxy clustering measurements~\cite{DESI_BAO_2024}. The large influx of \Lya forest data from the currently observing Dark Energy Spectroscopic Instrument \citep[DESI;][]{DESI:2016, DESI:2022}, paired with the recent development of the one-loop EFT power spectrum~\cite{Ivanov2024}, poses new opportunities but also challenges. First, informative priors on the large number of EFT bias parameters are essential for full-shape cosmological analyses of the data at hand. Second, akin to galaxy clustering an accurate estimate of the theoretical and systematics error budget is required~\cite{Chen:2024tfp}. Crucially, given that \Lya forest analyses rely on compressed statistics~\cite{duMasdesBourboux:2020pck, DESI_lya_2024}, this uncertainty is \textit{not} folded into the final BAO constraints.  
 
Our main conclusion is that the one-loop \Lya EFT power spectrum description fits the state-of-the-art hydrodynamical ACCEL$^2$ simulations in five redshift snapshots at the $2\%$ level on small scales and at the $10\%$ level on large scales. We obtain consistent results for two different box sizes in this suite: First, the one with side length $L=160 \Mpch$ and highest resolution down to $25 \kpch$ for our baseline results and second, the largest available box of size $L=640 \hinvMpc$ with an effective resolution of $100\, h^{-1}\mathrm{kpc}$ to access more linear modes. 

We use the resulting best-fit EFT bias parameters to theoretically derive the non-linear shift of the BAO signal in perturbation theory. We find a shift in the isotropic BAO scale and Alcock-Paczynski parameter at $\Delta \alphaiso = -0.28 \pm 0.09\, \%$ and $ \Delta \alphaap = 0.11 \pm 0.07\, \%$, respectively, when \Lya EFT parameters are sampled from the posterior of our fits to the 3D power spectrum.  
Our findings do not support the claim of a large, percent-level shift to $\alpha=0.9905 \pm 0.0027$ in redshift space 
made in Ref.~\cite{Sinigaglia:2024kub}.\footnote{Since we do not have a snapshot at the same redshift, we also compare the snapshot at redshift $z=2.6$, finding a similar result for the BAO shift, see Eqs.~\eqref{eq:BAOshift_auto_z20}-\eqref{eq:BAOshift_auto_z26_aiso}.}  For the shift of the BAO signal obtained from the \Lya--quasar cross-correlation using the bias relation obtained from~\cite{Abidi:2018eyd} in radial ($\apar$) and transverse ($\aperp$) direction, we obtain at $z=2.0$ the shifts $\Delta \alphaiso^{\times} = -0.11 \pm 0.04\, \%$ and $\Delta \alphaap^{\times} = -0.07 \pm 0.05 \%$.
These derived BAO shifts should be added as an error term to the covariance for BAO-based cosmological analyses, as is done in the case of galaxy BAO measurements, especially since the reconstruction of the \Lya BAO is not straightforward.

This paper has demonstrated the robustness of the one-loop power spectrum model in the EFT framework on hydrodynamical simulations and shows a promising path forward to perform cosmological analyses of the \Lya forest data from the currently observing DESI survey. Importantly, we show that the EFT framework 
can successfully describe 
data with different 
assumptions about the small-scale
physics in the large and small box 
ACCEL$^2$ simulations.
It is not obvious that 
a similar success can be achieved 
with emulators trained
on different types of simulations. 
This highlights the flexibility
of EFT and emphasizes one of its 
key advantages relevant for future 
\Lya data analyses.

A number of future surveys will capture spectra in both high and medium resolution modes, such as the WEAVE-QSO survey \citep{2016sf2a.conf..259P}, the Prime Focus Spectrograph \citep[PFS;][]{2022PFSGE}, and 4MOST \citep{2019Msngr.175....3D} posing interesting application possibilities which we leave to future work. We stress that future analyses including the cosmology dependence and the impact of the physics of the intergalactic medium are essential as well as understanding the physics of the EFT bias parameters. We leave a detailed analysis on simulations, including halos (or quasar catalogs) to future work. In a companion paper, we will present measurements at the field-level. This work paves the way to a number of future analyses such as full-shape analyses of the \Lya forest in Fourier space using the three-dimensional power spectrum~\cite{Font-Ribera:2018, deBelsunce:2024knf} with  field-level simulation-based 
priors~\cite{Ivanov:2024hgq,Ivanov:2024xgb}, or beyond BAO analyses using the configuration space correlation function~\cite{Cuceu:2021hlk}. 

\section*{Acknowledgments}
The authors thank Pat McDonald, Andreu Font-Ribera, Julien Guy, and Martin White for fruitful discussions and comments on the draft. 

SC acknowledges the support of the National Science Foundation at the Institute for Advanced Study. This research used resources of the National Energy Research Scientific Computing Center (NERSC), a U.S. Department of Energy Office of Science User Facility operated under Contract No.~DE–AC02–05CH11231. 

\appendix

\section{Nonlinear BAO shift}\label{app:BAO_shift}
In this appendix, we provide details on how we obtain an analytic expression for the non-linear BAO shift. To solve the integrals given in Sec.~\ref{sec:BAO_shift}, we adopt a spherical coordinate system $\hat{k} = \{ 0, 0, 1 \}\,, \hat{n} = \{ \sqrt{1 - \nu^2}, 0, \nu \}$ and $\hat{q} = \{ \cos(\phi) + \sqrt{1 - \mu^2}, \sin(\phi) + \sqrt{1 - \mu^2}, \mu \}$, following ~\cite{Chen:2024tfp}. The final analytic form for the non-linear BAO shift in \Lya forest clustering is obtained using \texttt{Mathematica}
\begin{widetext}
\begin{align} \label{eq:Pshift}
P_{\rm shift} &= -\frac{\sigma^2_d kP'_w(k)}{945}\Bigg[\left(b_1 - b_{\eta} f \mu^2\right) 
\left(423 b_1 + 630 b_2 - 504 b_{\mathcal{G}_2} + 56 b_{(KK)_\parallel} - 126 b_1 f - 126 b_{\delta \eta} f + \mu^2 \left(28 b_{(KK)_\parallel} \left(6 + 5 f\right)\right) \right. \nonumber\\
& \left. - \mu^2\left(36 b_{\Pi^{[2]}_\parallel} \left(31 + 7 f\right) + 9 f \left(12 b_1 + 70 b_2 - 98 b_{\delta \eta} - 56 b_{\mathcal{G}_2} - 31 b_{\eta} + 14 \left(-4 b_1 - 3 b_{\delta \eta} + 2 b_{\eta^2} + 3 b_{\eta}\right) f \right)\right) \right. \nonumber \\
& \left. - 3 f \left(28 b_{(KK)_\parallel} + 288 b_{\Pi^{[2]}_\parallel} + 3 f \left(7 b_1 - 70 b_{\delta \eta} + 28 b_{\eta^2} \left(2 + 3 f\right) + 8 b_{\eta} \left(-3 + 14 f\right)\right)\right) \mu^4 - 567 b_{\eta} f^3 \mu^6 \right) \Bigg]  \,,
\end{align}
\end{widetext}
where $\sigma^2_d$ is the mean-square overdensity on BAO scales (Eq.~\eqref{eqn:shift_term}) and $P'_w(k)=\frac{\partial P}{\partial\ln{k}}$ the response of the power spectrum. The BAO shift for the cross-correlation with quasars is
\begin{widetext}
\begin{align} \label{eq:Pshift_cross}
P_{\rm shift}^{\mathrm{cross}} &= \frac{\sigma^2_d kP'_w(k)}{945} \Bigg[ -423 b_1 b_{1}^q - 315 b_{1}^q b_2 - 315 b_1 b_{2}^q + 252 b_{1}^q b_{\mathcal{G}_2} + 252 b_1 b_{\mathcal{G}_2}^q - 28 b_{1}^q b_{(KK)_\parallel} 
+ 63 b_1 b_{1}^q f + 63 b_{1}^q b_{\delta \eta} f \nonumber \\
& - \left(351 b_1 f + 7 f \left(45 b_2 + 4 \left(-9 b_{\mathcal{G}_2} + b_{(KK)_\parallel}\right) + 9 \left(5 b_{2}^q - 4 b_{\mathcal{G}_2}^q\right) (b_1 - b_{\eta}) 
- 9 \left(2 b_1 + b_{\delta \eta}\right) f \right) \right) \nonumber \\
& - b_{1}^q \left(14 b_{(KK)_\parallel}(6 + 5 f) + 18 b_{\Pi^{[2]}}(31 + 7 f) + 9 f \left(61 b_1 + 35 b_2 - 49 b_{\delta \eta} 
- 28 b_{\mathcal{G}_2} - 39 b_{\eta} - 7 (5 b_1 + 3 b_{\delta \eta} - 2 b_{\eta^2} - 3 b_{\eta}) f \right) \right) \mu^2 \nonumber \\
& - f \left(14 b_{(KK)_\parallel}(6 + 3 b_{1}^q + 5 f) + 9 \left(2 (31 + 24 b_{1}^q) b_{\Pi^{[2]}} + 46 b_1 f 
+ 7 (6 b_1 b_{1}^q + 5 b_2 - 7 b_{\delta \eta} - 5 b_{1}^q b_{\delta \eta} + 4 b_{1}^q b_{\eta^2} - 4 b_{\mathcal{G}_2} + 2 b_{\Pi^{[2]}}) f \right. \right. \nonumber \\
& \phantom{-} - (31 + 67 b_{1}^q + 35 b_{2}^q - 28 b_{\mathcal{G}_2}^q) b_{\eta} f - 7 (6 b_1 + 3 b_{\delta \eta} - 2 b_{\eta^2} - 6 b_{1}^q b_{\eta^2} - 4 b_{\eta} - 9 b_{1}^q b_{\eta}) f^2) \mu^4 \nonumber \\
& - 3 f^2 \left(14 b_{(KK)_\parallel} + 144 b_{\Pi^{[2]}} + 3 f \left(35 b_1 - 35 b_{\delta \eta} + 28 b_{\eta^2} - 52 b_{\eta} - 70 b_{1}^q b_{\eta} + 42 b_{\eta^2} f + 70 b_{\eta} f \right) \right) \mu^6 \nonumber \\
& + 567 b_{\eta} f^4 \mu^8 \Bigg] \,.
\end{align}
\end{widetext}

\bibliographystyle{JHEP}
\bibliography{short.bib, references}

\providecommand{\href}[2]{#2}\begingroup\raggedright\begin{thebibliography}{100}

\bibitem{McDonald06}
P.~{McDonald}, U.~{Seljak}, S.~{Burles}, D.~J. {Schlegel}, D.~H. {Weinberg}, R.~{Cen} et~al., \emph{{The Ly{\ensuremath{\alpha}} Forest Power Spectrum from the Sloan Digital Sky Survey}}, \href{https://doi.org/10.1086/444361}{\emph{\apjs} {\bfseries 163} (2006) 80} [\href{https://arxiv.org/abs/astro-ph/0405013}{{\ttfamily astro-ph/0405013}}].

\bibitem{PYB13}
N.~{Palanque-Delabrouille}, C.~{Y{\`e}che}, A.~{Borde}, J.-M. {Le Goff}, G.~{Rossi}, M.~{Viel} et~al., \emph{{The one-dimensional Ly{\ensuremath{\alpha}} forest power spectrum from BOSS}}, \href{https://doi.org/10.1051/0004-6361/201322130}{\emph{\aap} {\bfseries 559} (2013) A85} [\href{https://arxiv.org/abs/1306.5896}{{\ttfamily 1306.5896}}].

\bibitem{DESI:2023xwh}
{\scshape DESI} collaboration, \emph{{The Dark Energy Spectroscopic Instrument: one-dimensional power spectrum from first Ly~\ensuremath{\alpha} forest samples with Fast Fourier Transform}}, \href{https://doi.org/10.1093/mnras/stad3008}{\emph{Mon. Not. Roy. Astron. Soc.} {\bfseries 526} (2023) 5118} [\href{https://arxiv.org/abs/2306.06311}{{\ttfamily 2306.06311}}].

\bibitem{Karacayli:2023afs}
N.~G. Kara\c{c}ayl\i{} et~al., \emph{{Optimal 1D Ly$\alpha$ Forest Power Spectrum Estimation -- III. DESI early data}}, \href{https://doi.org/10.1093/mnras/stae171}{\emph{Mon. Not. Roy. Astron. Soc.} {\bfseries 528} (2024) 3941} [\href{https://arxiv.org/abs/2306.06316}{{\ttfamily 2306.06316}}].

\bibitem{Seljak:2005}
U.~{Seljak}, A.~{Makarov}, P.~{McDonald}, S.~F. {Anderson}, N.~A. {Bahcall}, J.~{Brinkmann} et~al., \emph{{Cosmological parameter analysis including SDSS Ly{\ensuremath{\alpha}} forest and galaxy bias: Constraints on the primordial spectrum of fluctuations, neutrino mass, and dark energy}}, \href{https://doi.org/10.1103/PhysRevD.71.103515}{\emph{\prd} {\bfseries 71} (2005) 103515} [\href{https://arxiv.org/abs/astro-ph/0407372}{{\ttfamily astro-ph/0407372}}].

\bibitem{Viel:2010}
M.~{Viel}, M.~G. {Haehnelt} and V.~{Springel}, \emph{{The effect of neutrinos on the matter distribution as probed by the intergalactic medium}}, \href{https://doi.org/10.1088/1475-7516/2010/06/015}{\emph{\jcap} {\bfseries 2010} (2010) 015} [\href{https://arxiv.org/abs/1003.2422}{{\ttfamily 1003.2422}}].

\bibitem{Palanque2020}
N.~{Palanque-Delabrouille}, C.~{Y{\`e}che}, N.~{Sch{\"o}neberg}, J.~{Lesgourgues}, M.~{Walther}, S.~{Chabanier} et~al., \emph{{Hints, neutrino bounds, and WDM constraints from SDSS DR14 Lyman-{\ensuremath{\alpha}} and Planck full-survey data}}, \href{https://doi.org/10.1088/1475-7516/2020/04/038}{\emph{\jcap} {\bfseries 2020} (2020) 038} [\href{https://arxiv.org/abs/1911.09073}{{\ttfamily 1911.09073}}].

\bibitem{Afshordi:2003}
N.~{Afshordi}, P.~{McDonald} and D.~N. {Spergel}, \emph{{Primordial Black Holes as Dark Matter: The Power Spectrum and Evaporation of Early Structures}}, \href{https://doi.org/10.1086/378763}{\emph{\apjl} {\bfseries 594} (2003) L71} [\href{https://arxiv.org/abs/astro-ph/0302035}{{\ttfamily astro-ph/0302035}}].

\bibitem{Murgia:2019}
R.~Murgia, G.~Scelfo, M.~Viel and A.~Raccanelli, \emph{{Lyman-\ensuremath{\alpha} Forest Constraints on Primordial Black Holes as Dark Matter}}, \href{https://doi.org/10.1103/PhysRevLett.123.071102}{\emph{Phys. Rev. Lett.} {\bfseries 123} (2019) 071102} [\href{https://arxiv.org/abs/1903.10509}{{\ttfamily 1903.10509}}].

\bibitem{Viel:2013}
M.~{Viel}, G.~D. {Becker}, J.~S. {Bolton} and M.~G. {Haehnelt}, \emph{{Warm dark matter as a solution to the small scale crisis: New constraints from high redshift Lyman-{\ensuremath{\alpha}} forest data}}, \href{https://doi.org/10.1103/PhysRevD.88.043502}{\emph{\prd} {\bfseries 88} (2013) 043502} [\href{https://arxiv.org/abs/1306.2314}{{\ttfamily 1306.2314}}].

\bibitem{Baur:2016}
J.~{Baur}, N.~{Palanque-Delabrouille}, C.~{Y{\`e}che}, C.~{Magneville} and M.~{Viel}, \emph{{Lyman-alpha forests cool warm dark matter}}, \href{https://doi.org/10.1088/1475-7516/2016/08/012}{\emph{\jcap} {\bfseries 2016} (2016) 012} [\href{https://arxiv.org/abs/1512.01981}{{\ttfamily 1512.01981}}].

\bibitem{Irsic17}
V.~{Ir{\v s}i{\v c}}, M.~{Viel}, M.~G. {Haehnelt}, J.~S. {Bolton}, S.~{Cristiani}, G.~D. {Becker} et~al., \emph{{New Constraints on the free-streaming of warm dark matter from intermediate and small scale Lyman-$\alpha$ forest data}}, {\emph{ArXiv e-prints} (2017) } [\href{https://arxiv.org/abs/1702.01764}{{\ttfamily 1702.01764}}].

\bibitem{Kobayashi:2017}
T.~{Kobayashi}, R.~{Murgia}, A.~{De Simone}, V.~{Ir{\v{s}}i{\v{c}}} and M.~{Viel}, \emph{{Lyman-{\ensuremath{\alpha}} constraints on ultralight scalar dark matter: Implications for the early and late universe}}, \href{https://doi.org/10.1103/PhysRevD.96.123514}{\emph{\prd} {\bfseries 96} (2017) 123514} [\href{https://arxiv.org/abs/1708.00015}{{\ttfamily 1708.00015}}].

\bibitem{Armengaud:2017}
E.~{Armengaud}, N.~{Palanque-Delabrouille}, C.~{Y{\`e}che}, D.~J.~E. {Marsh} and J.~{Baur}, \emph{{Constraining the mass of light bosonic dark matter using SDSS Lyman-{\ensuremath{\alpha}} forest}}, \href{https://doi.org/10.1093/mnras/stx1870}{\emph{\mnras} {\bfseries 471} (2017) 4606} [\href{https://arxiv.org/abs/1703.09126}{{\ttfamily 1703.09126}}].

\bibitem{Murgia:2018}
R.~{Murgia}, V.~{Ir{\v{s}}i{\v{c}}} and M.~{Viel}, \emph{{Novel constraints on noncold, nonthermal dark matter from Lyman-{\ensuremath{\alpha}} forest data}}, \href{https://doi.org/10.1103/PhysRevD.98.083540}{\emph{\prd} {\bfseries 98} (2018) 083540} [\href{https://arxiv.org/abs/1806.08371}{{\ttfamily 1806.08371}}].

\bibitem{Garzilli:2019}
A.~{Garzilli}, A.~{Magalich}, T.~{Theuns}, C.~S. {Frenk}, C.~{Weniger}, O.~{Ruchayskiy} et~al., \emph{{The Lyman-{\ensuremath{\alpha}} forest as a diagnostic of the nature of the dark matter}}, \href{https://doi.org/10.1093/mnras/stz2188}{\emph{\mnras} {\bfseries 489} (2019) 3456} [\href{https://arxiv.org/abs/1809.06585}{{\ttfamily 1809.06585}}].

\bibitem{Irsic:2020}
V.~{Ir{\v{s}}i{\v{c}}}, H.~{Xiao} and M.~{McQuinn}, \emph{{Early structure formation constraints on the ultralight axion in the postinflation scenario}}, \href{https://doi.org/10.1103/PhysRevD.101.123518}{\emph{\prd} {\bfseries 101} (2020) 123518} [\href{https://arxiv.org/abs/1911.11150}{{\ttfamily 1911.11150}}].

\bibitem{Rogers:2022}
K.~K. {Rogers}, C.~{Dvorkin} and H.~V. {Peiris}, \emph{{Limits on the Light Dark Matter-Proton Cross Section from Cosmic Large-Scale Structure}}, \href{https://doi.org/10.1103/PhysRevLett.128.171301}{\emph{\prl} {\bfseries 128} (2022) 171301} [\href{https://arxiv.org/abs/2111.10386}{{\ttfamily 2111.10386}}].

\bibitem{Villasenor:2023}
B.~{Villasenor}, B.~{Robertson}, P.~{Madau} and E.~{Schneider}, \emph{{New constraints on warm dark matter from the Lyman-{\ensuremath{\alpha}} forest power spectrum}}, \href{https://doi.org/10.1103/PhysRevD.108.023502}{\emph{\prd} {\bfseries 108} (2023) 023502} [\href{https://arxiv.org/abs/2209.14220}{{\ttfamily 2209.14220}}].

\bibitem{Irsic:2023}
V.~{Ir{\v{s}}i{\v{c}}}, M.~{Viel}, M.~G. {Haehnelt}, J.~S. {Bolton}, M.~{Molaro}, E.~{Puchwein} et~al., \emph{{Unveiling Dark Matter free-streaming at the smallest scales with high redshift Lyman-alpha forest}}, \href{https://doi.org/10.48550/arXiv.2309.04533}{\emph{arXiv e-prints} (2023) arXiv:2309.04533} [\href{https://arxiv.org/abs/2309.04533}{{\ttfamily 2309.04533}}].

\bibitem{2023PhRvL.131t1001G}
S.~{Goldstein}, J.~C. {Hill}, V.~{Ir{\v{s}}i{\v{c}}} and B.~D. {Sherwin}, \emph{{Canonical Hubble-Tension-Resolving Early Dark Energy Cosmologies Are Inconsistent with the Lyman-{\ensuremath{\alpha}} Forest}}, \href{https://doi.org/10.1103/PhysRevLett.131.201001}{\emph{\prl} {\bfseries 131} (2023) 201001} [\href{https://arxiv.org/abs/2303.00746}{{\ttfamily 2303.00746}}].

\bibitem{Zaldarriaga:2002}
M.~{Zaldarriaga}, \emph{{Searching for Fluctuations in the Intergalactic Medium Temperature Using the Ly{\ensuremath{\alpha}} Forest}}, \href{https://doi.org/10.1086/324212}{\emph{\apj} {\bfseries 564} (2002) 153} [\href{https://arxiv.org/abs/astro-ph/0102205}{{\ttfamily astro-ph/0102205}}].

\bibitem{Meiksin:2009}
A.~A. {Meiksin}, \emph{{The physics of the intergalactic medium}}, \href{https://doi.org/10.1103/RevModPhys.81.1405}{\emph{Reviews of Modern Physics} {\bfseries 81} (2009) 1405} [\href{https://arxiv.org/abs/0711.3358}{{\ttfamily 0711.3358}}].

\bibitem{McQuinn:2016}
M.~{McQuinn}, \emph{{The Evolution of the Intergalactic Medium}}, \href{https://doi.org/10.1146/annurev-astro-082214-122355}{\emph{\araa} {\bfseries 54} (2016) 313} [\href{https://arxiv.org/abs/1512.00086}{{\ttfamily 1512.00086}}].

\bibitem{Viel:2006}
M.~{Viel}, J.~{Lesgourgues}, M.~G. {Haehnelt}, S.~{Matarrese} and A.~{Riotto}, \emph{{Can Sterile Neutrinos Be Ruled Out as Warm Dark Matter Candidates?}}, \href{https://doi.org/10.1103/PhysRevLett.97.071301}{\emph{\prl} {\bfseries 97} (2006) 071301} [\href{https://arxiv.org/abs/astro-ph/0605706}{{\ttfamily astro-ph/0605706}}].

\bibitem{Walther:2019}
M.~{Walther}, J.~{O{\~n}orbe}, J.~F. {Hennawi} and Z.~{Luki{\'c}}, \emph{{New Constraints on IGM Thermal Evolution from the Ly{\ensuremath{\alpha}} Forest Power Spectrum}}, \href{https://doi.org/10.3847/1538-4357/aafad1}{\emph{\apj} {\bfseries 872} (2019) 13} [\href{https://arxiv.org/abs/1808.04367}{{\ttfamily 1808.04367}}].

\bibitem{Bolton:2008}
J.~S. {Bolton}, M.~{Viel}, T.~S. {Kim}, M.~G. {Haehnelt} and R.~F. {Carswell}, \emph{{Possible evidence for an inverted temperature-density relation in the intergalactic medium from the flux distribution of the Ly{\ensuremath{\alpha}} forest}}, \href{https://doi.org/10.1111/j.1365-2966.2008.13114.x}{\emph{\mnras} {\bfseries 386} (2008) 1131} [\href{https://arxiv.org/abs/0711.2064}{{\ttfamily 0711.2064}}].

\bibitem{Garzilli:2012}
A.~{Garzilli}, J.~S. {Bolton}, T.~S. {Kim}, S.~{Leach} and M.~{Viel}, \emph{{The intergalactic medium thermal history at redshift z = 1.7-3.2 from the Ly{\ensuremath{\alpha}} forest: a comparison of measurements using wavelets and the flux distribution}}, \href{https://doi.org/10.1111/j.1365-2966.2012.21223.x}{\emph{\mnras} {\bfseries 424} (2012) 1723} [\href{https://arxiv.org/abs/1202.3577}{{\ttfamily 1202.3577}}].

\bibitem{Gaikwad:2019}
P.~{Gaikwad}, R.~{Srianand}, V.~{Khaire} and T.~R. {Choudhury}, \emph{{Effect of non-equilibrium ionization on derived physical conditions of the high-z intergalactic medium}}, \href{https://doi.org/10.1093/mnras/stz2692}{\emph{\mnras} {\bfseries 490} (2019) 1588} [\href{https://arxiv.org/abs/1812.01016}{{\ttfamily 1812.01016}}].

\bibitem{Boera:2019}
E.~{Boera}, G.~D. {Becker}, J.~S. {Bolton} and F.~{Nasir}, \emph{{Revealing Reionization with the Thermal History of the Intergalactic Medium: New Constraints from the Ly{\ensuremath{\alpha}} Flux Power Spectrum}}, \href{https://doi.org/10.3847/1538-4357/aafee4}{\emph{\apj} {\bfseries 872} (2019) 101} [\href{https://arxiv.org/abs/1809.06980}{{\ttfamily 1809.06980}}].

\bibitem{Gaikwad:2021}
P.~{Gaikwad}, R.~{Srianand}, M.~G. {Haehnelt} and T.~R. {Choudhury}, \emph{{A consistent and robust measurement of the thermal state of the IGM at 2 {\ensuremath{\leq}} z {\ensuremath{\leq}} 4 from a large sample of Ly {\ensuremath{\alpha}} forest spectra: evidence for late and rapid He II reionization}}, \href{https://doi.org/10.1093/mnras/stab2017}{\emph{\mnras} {\bfseries 506} (2021) 4389} [\href{https://arxiv.org/abs/2009.00016}{{\ttfamily 2009.00016}}].

\bibitem{Wilson:2022}
B.~{Wilson}, V.~{Ir{\v{s}}i{\v{c}}} and M.~{McQuinn}, \emph{{A measurement of the Ly {\ensuremath{\beta}} forest power spectrum and its cross with the Ly {\ensuremath{\alpha}} forest in X-Shooter XQ-100}}, \href{https://doi.org/10.1093/mnras/stab3017}{\emph{\mnras} {\bfseries 509} (2022) 2423} [\href{https://arxiv.org/abs/2106.04837}{{\ttfamily 2106.04837}}].

\bibitem{Villasenor:2022}
B.~{Villasenor}, B.~{Robertson}, P.~{Madau} and E.~{Schneider}, \emph{{Inferring the Thermal History of the Intergalactic Medium from the Properties of the Hydrogen and Helium Ly{\ensuremath{\alpha}} Forest}}, \href{https://doi.org/10.3847/1538-4357/ac704e}{\emph{\apj} {\bfseries 933} (2022) 59} [\href{https://arxiv.org/abs/2111.00019}{{\ttfamily 2111.00019}}].

\bibitem{McDonald:2007}
P.~{McDonald} and D.~J. {Eisenstein}, \emph{{Dark energy and curvature from a future baryonic acoustic oscillation survey using the Lyman-{\ensuremath{\alpha}} forest}}, \href{https://doi.org/10.1103/PhysRevD.76.063009}{\emph{\prd} {\bfseries 76} (2007) 063009} [\href{https://arxiv.org/abs/astro-ph/0607122}{{\ttfamily astro-ph/0607122}}].

\bibitem{Slosar2013}
A.~{Slosar}, V.~{Ir{\v{s}}i{\v{c}}}, D.~{Kirkby}, S.~{Bailey}, N.~G. {Busca}, T.~{Delubac} et~al., \emph{{Measurement of baryon acoustic oscillations in the Lyman-{\ensuremath{\alpha}} forest fluctuations in BOSS data release 9}}, \href{https://doi.org/10.1088/1475-7516/2013/04/026}{\emph{\jcap} {\bfseries 2013} (2013) 026} [\href{https://arxiv.org/abs/1301.3459}{{\ttfamily 1301.3459}}].

\bibitem{Busca:2013}
N.~G. {Busca}, T.~{Delubac}, J.~{Rich}, S.~{Bailey}, A.~{Font-Ribera}, D.~{Kirkby} et~al., \emph{{Baryon acoustic oscillations in the Ly{\ensuremath{\alpha}} forest of BOSS quasars}}, \href{https://doi.org/10.1051/0004-6361/201220724}{\emph{\aap} {\bfseries 552} (2013) A96} [\href{https://arxiv.org/abs/1211.2616}{{\ttfamily 1211.2616}}].

\bibitem{duMasdesBourboux:2020pck}
H.~du~Mas~des Bourboux et~al., \emph{{The Completed SDSS-IV Extended Baryon Oscillation Spectroscopic Survey: Baryon Acoustic Oscillations with Ly\ensuremath{\alpha} Forests}}, \href{https://doi.org/10.3847/1538-4357/abb085}{\emph{Astrophys. J.} {\bfseries 901} (2020) 153} [\href{https://arxiv.org/abs/2007.08995}{{\ttfamily 2007.08995}}].

\bibitem{DESI_lya_2024}
{DESI Collaboration}, A.~G. {Adame}, J.~{Aguilar}, S.~{Ahlen}, S.~{Alam}, D.~M. {Alexander} et~al., \emph{{DESI 2024 IV: Baryon Acoustic Oscillations from the Lyman Alpha Forest}}, \href{https://doi.org/10.48550/arXiv.2404.03001}{\emph{arXiv e-prints} (2024) arXiv:2404.03001} [\href{https://arxiv.org/abs/2404.03001}{{\ttfamily 2404.03001}}].

\bibitem{eBOSS:2015jyv}
{\scshape eBOSS} collaboration, \emph{{The SDSS-IV extended Baryon Oscillation Spectroscopic Survey: Overview and Early Data}}, \href{https://doi.org/10.3847/0004-6256/151/2/44}{\emph{Astron. J.} {\bfseries 151} (2016) 44} [\href{https://arxiv.org/abs/1508.04473}{{\ttfamily 1508.04473}}].

\bibitem{DESI:2016fyo}
{\scshape DESI} collaboration, \emph{{The DESI Experiment Part I: Science,Targeting, and Survey Design}},  \href{https://arxiv.org/abs/1611.00036}{{\ttfamily 1611.00036}}.

\bibitem{DESI:2022}
B.~{Abareshi}, J.~{Aguilar}, S.~{Ahlen}, S.~{Alam}, D.~M. {Alexander}, R.~{Alfarsy} et~al., \emph{{Overview of the Instrumentation for the Dark Energy Spectroscopic Instrument}}, {\emph{arXiv e-prints} (2022) arXiv:2205.10939} [\href{https://arxiv.org/abs/2205.10939}{{\ttfamily 2205.10939}}].

\bibitem{Baumann:2010tm}
D.~Baumann, A.~Nicolis, L.~Senatore and M.~Zaldarriaga, \emph{{Cosmological Non-Linearities as an Effective Fluid}}, \href{https://doi.org/10.1088/1475-7516/2012/07/051}{\emph{JCAP} {\bfseries 1207} (2012) 051} [\href{https://arxiv.org/abs/1004.2488}{{\ttfamily 1004.2488}}].

\bibitem{Carrasco:2013mua}
J.~J.~M. Carrasco, S.~Foreman, D.~Green and L.~Senatore, \emph{{The Effective Field Theory of Large Scale Structures at Two Loops}}, \href{https://doi.org/10.1088/1475-7516/2014/07/057}{\emph{JCAP} {\bfseries 07} (2014) 057} [\href{https://arxiv.org/abs/1310.0464}{{\ttfamily 1310.0464}}].

\bibitem{Ivanov:2022mrd}
M.~M. Ivanov, \emph{{Effective Field Theory for Large Scale Structure}},  \href{https://arxiv.org/abs/2212.08488}{{\ttfamily 2212.08488}}.

\bibitem{McDonald:2009dh}
P.~McDonald and A.~Roy, \emph{{Clustering of dark matter tracers: generalizing bias for the coming era of precision LSS}}, \href{https://doi.org/10.1088/1475-7516/2009/08/020}{\emph{JCAP} {\bfseries 0908} (2009) 020} [\href{https://arxiv.org/abs/0902.0991}{{\ttfamily 0902.0991}}].

\bibitem{Givans:2020sez}
J.~J. Givans and C.~M. Hirata, \emph{{Redshift-space streaming velocity effects on the Lyman-$\alpha$ forest baryon acoustic oscillation scale}}, \href{https://doi.org/10.1103/PhysRevD.102.023515}{\emph{Phys. Rev. D} {\bfseries 102} (2020) 023515} [\href{https://arxiv.org/abs/2002.12296}{{\ttfamily 2002.12296}}].

\bibitem{Desjacques:2018pfv}
V.~Desjacques, D.~Jeong and F.~Schmidt, \emph{{The Galaxy Power Spectrum and Bispectrum in Redshift Space}}, \href{https://doi.org/10.1088/1475-7516/2018/12/035}{\emph{JCAP} {\bfseries 1812} (2018) 035} [\href{https://arxiv.org/abs/1806.04015}{{\ttfamily 1806.04015}}].

\bibitem{Chen:2021rnb}
S.-F. Chen, Z.~Vlah and M.~White, \emph{{The Ly$\alpha$ forest flux correlation function: a perturbation theory perspective}}, \href{https://doi.org/10.1088/1475-7516/2021/05/053}{\emph{JCAP} {\bfseries 05} (2021) 053} [\href{https://arxiv.org/abs/2103.13498}{{\ttfamily 2103.13498}}].

\bibitem{Ivanov2024}
M.~M. {Ivanov}, \emph{{Lyman alpha forest power spectrum in effective field theory}}, \href{https://doi.org/10.1103/PhysRevD.109.023507}{\emph{\prd} {\bfseries 109} (2024) 023507} [\href{https://arxiv.org/abs/2309.10133}{{\ttfamily 2309.10133}}].

\bibitem{Ivanov:2024jtl}
M.~M. Ivanov, M.~W. Toomey and N.~G. Kara\c{c}ayl\i{}, \emph{{Fundamental physics with the Lyman-alpha forest: constraints on the growth of structure and neutrino masses from SDSS with effective field theory}},  \href{https://arxiv.org/abs/2405.13208}{{\ttfamily 2405.13208}}.

\bibitem{Gerardi:2022ncj}
F.~Gerardi, A.~Cuceu, A.~Font-Ribera, B.~Joachimi and P.~Lemos, \emph{{Direct cosmological inference from three-dimensional correlations of the Lyman \ensuremath{\alpha} forest}}, \href{https://doi.org/10.1093/mnras/stac3257}{\emph{Mon. Not. Roy. Astron. Soc.} {\bfseries 518} (2022) 2567} [\href{https://arxiv.org/abs/2209.11263}{{\ttfamily 2209.11263}}].

\bibitem{Cuceu:2021hlk}
A.~Cuceu, A.~Font-Ribera, B.~Joachimi and S.~Nadathur, \emph{{Cosmology beyond BAO from the 3D distribution of the Lyman-\ensuremath{\alpha} forest}}, \href{https://doi.org/10.1093/mnras/stab1999}{\emph{Mon. Not. Roy. Astron. Soc.} {\bfseries 506} (2021) 5439} [\href{https://arxiv.org/abs/2103.14075}{{\ttfamily 2103.14075}}].

\bibitem{Font-Ribera:2018}
A.~{Font-Ribera}, P.~{McDonald} and A.~{Slosar}, \emph{{How to estimate the 3D power spectrum of the Lyman-{\ensuremath{\alpha}} forest}}, \href{https://doi.org/10.1088/1475-7516/2018/01/003}{\emph{\jcap} {\bfseries 2018} (2018) 003} [\href{https://arxiv.org/abs/1710.11036}{{\ttfamily 1710.11036}}].

\bibitem{Abdul-Karim:2023sgf}
M.~L. Abdul-Karim, E.~Armengaud, G.~Mention, S.~Chabanier, C.~Ravoux and Z.~Luki\'c, \emph{{Measurement of the small-scale 3D Lyman-\ensuremath{\alpha} forest power spectrum}}, \href{https://doi.org/10.1088/1475-7516/2024/05/088}{\emph{JCAP} {\bfseries 05} (2024) 088} [\href{https://arxiv.org/abs/2310.09116}{{\ttfamily 2310.09116}}].

\bibitem{deBelsunce:2024knf}
R.~de~Belsunce, O.~H.~E. Philcox, V.~Irsic, P.~McDonald, J.~Guy and N.~Palanque-Delabrouille, \emph{{The 3D Lyman-\ensuremath{\alpha} forest power spectrum from eBOSS DR16}}, \href{https://doi.org/10.1093/mnras/stae2035}{\emph{Mon. Not. Roy. Astron. Soc.} {\bfseries 533} (2024) 3756} [\href{https://arxiv.org/abs/2403.08241}{{\ttfamily 2403.08241}}].

\bibitem{Horowitz:2024nny}
B.~Horowitz, R.~de~Belsunce and Z.~Lukic, \emph{{Maximum A Posteriori Ly-alpha Estimator (MAPLE): Band-power and covariance estimation of the 3D Ly-alpha forest power spectrum}},  \href{https://arxiv.org/abs/2403.17294}{{\ttfamily 2403.17294}}.

\bibitem{Ivanov:2019pdj}
M.~M. Ivanov, M.~Simonovi\'c and M.~Zaldarriaga, \emph{{Cosmological Parameters from the BOSS Galaxy Power Spectrum}}, \href{https://doi.org/10.1088/1475-7516/2020/05/042}{\emph{JCAP} {\bfseries 05} (2020) 042} [\href{https://arxiv.org/abs/1909.05277}{{\ttfamily 1909.05277}}].

\bibitem{DAmico:2019fhj}
G.~D'Amico, J.~Gleyzes, N.~Kokron, D.~Markovic, L.~Senatore, P.~Zhang et~al., \emph{{The Cosmological Analysis of the SDSS/BOSS data from the Effective Field Theory of Large-Scale Structure}},  \href{https://arxiv.org/abs/1909.05271}{{\ttfamily 1909.05271}}.

\bibitem{Chen:2021wdi}
S.-F. Chen, Z.~Vlah and M.~White, \emph{{A new analysis of galaxy 2-point functions in the BOSS survey, including full-shape information and post-reconstruction BAO}}, \href{https://doi.org/10.1088/1475-7516/2022/02/008}{\emph{JCAP} {\bfseries 02} (2022) 008} [\href{https://arxiv.org/abs/2110.05530}{{\ttfamily 2110.05530}}].

\bibitem{Philcox:2021kcw}
O.~H.~E. Philcox and M.~M. Ivanov, \emph{{BOSS DR12 full-shape cosmology: \ensuremath{\Lambda}CDM constraints from the large-scale galaxy power spectrum and bispectrum monopole}}, \href{https://doi.org/10.1103/PhysRevD.105.043517}{\emph{Phys. Rev. D} {\bfseries 105} (2022) 043517} [\href{https://arxiv.org/abs/2112.04515}{{\ttfamily 2112.04515}}].

\bibitem{Chen:2022jzq}
S.-F. Chen, M.~White, J.~DeRose and N.~Kokron, \emph{{Cosmological analysis of three-dimensional BOSS galaxy clustering and Planck CMB lensing cross correlations via Lagrangian perturbation theory}}, \href{https://doi.org/10.1088/1475-7516/2022/07/041}{\emph{JCAP} {\bfseries 07} (2022) 041} [\href{https://arxiv.org/abs/2204.10392}{{\ttfamily 2204.10392}}].

\bibitem{Chen:2024vuf}
S.-F. Chen, M.~M. Ivanov, O.~H.~E. Philcox and L.~Wenzl, \emph{{Suppression without Thawing: Constraining Structure Formation and Dark Energy with Galaxy Clustering}},  \href{https://arxiv.org/abs/2406.13388}{{\ttfamily 2406.13388}}.

\bibitem{Chabanier:2024knr}
S.~Chabanier, C.~Ravoux, L.~Latrille, J.~Sexton, E.~Armengaud, J.~Bautista et~al., \emph{{The ACCEL$^2$ project: simulating Lyman-$\alpha$ forest in large-volume hydrodynamical simulations}},  \href{https://arxiv.org/abs/2407.04473}{{\ttfamily 2407.04473}}.

\bibitem{McDonald:2001fe}
P.~McDonald, \emph{{Toward a measurement of the cosmological geometry at Z 2: predicting lyman-alpha forest correlation in three dimensions, and the potential of future data sets}}, \href{https://doi.org/10.1086/345945}{\emph{Astrophys. J.} {\bfseries 585} (2003) 34} [\href{https://arxiv.org/abs/astro-ph/0108064}{{\ttfamily astro-ph/0108064}}].

\bibitem{Arinyo-i-Prats:2015vqa}
A.~Arinyo-i Prats, J.~Miralda-Escud\'e, M.~Viel and R.~Cen, \emph{{The Non-Linear Power Spectrum of the Lyman Alpha Forest}}, \href{https://doi.org/10.1088/1475-7516/2015/12/017}{\emph{JCAP} {\bfseries 12} (2015) 017} [\href{https://arxiv.org/abs/1506.04519}{{\ttfamily 1506.04519}}].

\bibitem{Givans2022}
J.~J. {Givans}, A.~{Font-Ribera}, A.~{Slosar}, L.~{Seeyave}, C.~{Pedersen}, K.~K. {Rogers} et~al., \emph{{Non-linearities in the Lyman-{\ensuremath{\alpha}} forest and in its cross-correlation with dark matter halos}}, \href{https://doi.org/10.1088/1475-7516/2022/09/070}{\emph{\jcap} {\bfseries 2022} (2022) 070} [\href{https://arxiv.org/abs/2205.00962}{{\ttfamily 2205.00962}}].

\bibitem{DESI_BAO_2024}
{DESI Collaboration}, A.~G. {Adame}, J.~{Aguilar}, S.~{Ahlen}, S.~{Alam}, D.~M. {Alexander} et~al., \emph{{DESI 2024 VI: Cosmological Constraints from the Measurements of Baryon Acoustic Oscillations}}, \href{https://doi.org/10.48550/arXiv.2404.03002}{\emph{arXiv e-prints} (2024) arXiv:2404.03002} [\href{https://arxiv.org/abs/2404.03002}{{\ttfamily 2404.03002}}].

\bibitem{SDSS:2005xqv}
{\scshape SDSS} collaboration, \emph{{Detection of the Baryon Acoustic Peak in the Large-Scale Correlation Function of SDSS Luminous Red Galaxies}}, \href{https://doi.org/10.1086/466512}{\emph{Astrophys. J.} {\bfseries 633} (2005) 560} [\href{https://arxiv.org/abs/astro-ph/0501171}{{\ttfamily astro-ph/0501171}}].

\bibitem{Cole:2005sx}
{\scshape 2dFGRS} collaboration, \emph{{The 2dF Galaxy Redshift Survey: Power-spectrum analysis of the final dataset and cosmological implications}}, \href{https://doi.org/10.1111/j.1365-2966.2005.09318.x}{\emph{Mon. Not. Roy. Astron. Soc.} {\bfseries 362} (2005) 505} [\href{https://arxiv.org/abs/astro-ph/0501174}{{\ttfamily astro-ph/0501174}}].

\bibitem{Planck2018}
{Planck Collaboration}, N.~{Aghanim}, Y.~{Akrami}, F.~{Arroja}, M.~{Ashdown}, J.~{Aumont} et~al., \emph{{Planck 2018 results. I. Overview and the cosmological legacy of Planck}}, \href{https://doi.org/10.1051/0004-6361/201833880}{\emph{\aap} {\bfseries 641} (2020) A1} [\href{https://arxiv.org/abs/1807.06205}{{\ttfamily 1807.06205}}].

\bibitem{Chen:2024tfp}
S.-F. Chen et~al., \emph{{Baryon Acoustic Oscillation Theory and Modelling Systematics for the DESI 2024 results}},  \href{https://arxiv.org/abs/2402.14070}{{\ttfamily 2402.14070}}.

\bibitem{Sinigaglia:2024kub}
F.~Sinigaglia, F.-S. Kitaura, K.~Nagamine and Y.~Oku, \emph{{The Negative Baryon Acoustic Oscillation Shift in the Ly\ensuremath{\alpha} Forest from Cosmological Simulations}}, \href{https://doi.org/10.3847/2041-8213/ad66bf}{\emph{Astrophys. J. Lett.} {\bfseries 971} (2024) L22} [\href{https://arxiv.org/abs/2407.03918}{{\ttfamily 2407.03918}}].

\bibitem{Farr20}
J.~{Farr}, A.~{Font-Ribera}, H.~{du Mas des Bourboux}, A.~{Mu{\~n}oz-Guti{\'e}rrez}, F.~J. {S{\'a}nchez}, A.~{Pontzen} et~al., \emph{{LyaCoLoRe: synthetic datasets for current and future Lyman-{\ensuremath{\alpha}} forest BAO surveys}}, \href{https://doi.org/10.1088/1475-7516/2020/03/068}{\emph{\jcap} {\bfseries 2020} (2020) 068} [\href{https://arxiv.org/abs/1912.02763}{{\ttfamily 1912.02763}}].

\bibitem{Font-Ribera:2013fha}
A.~Font-Ribera et~al., \emph{{The large-scale Quasar-Lyman \textbackslash{}alpha\textbackslash{} Forest Cross-Correlation from BOSS}}, \href{https://doi.org/10.1088/1475-7516/2013/05/018}{\emph{JCAP} {\bfseries 05} (2013) 018} [\href{https://arxiv.org/abs/1303.1937}{{\ttfamily 1303.1937}}].

\bibitem{Chudaykin:2022nru}
A.~Chudaykin and M.~M. Ivanov, \emph{{Cosmological constraints from the power spectrum of eBOSS quasars}},  \href{https://arxiv.org/abs/2210.17044}{{\ttfamily 2210.17044}}.

\bibitem{Abidi:2018eyd}
M.~M. Abidi and T.~Baldauf, \emph{{Cubic Halo Bias in Eulerian and Lagrangian Space}}, \href{https://doi.org/10.1088/1475-7516/2018/07/029}{\emph{JCAP} {\bfseries 1807} (2018) 029} [\href{https://arxiv.org/abs/1802.07622}{{\ttfamily 1802.07622}}].

\bibitem{Planck:2015fie}
{\scshape Planck} collaboration, \emph{{Planck 2015 results. XIII. Cosmological parameters}}, \href{https://doi.org/10.1051/0004-6361/201525830}{\emph{Astron. Astrophys.} {\bfseries 594} (2016) A13} [\href{https://arxiv.org/abs/1502.01589}{{\ttfamily 1502.01589}}].

\bibitem{Almgren:2013}
A.~S. {Almgren}, J.~B. {Bell}, M.~J. {Lijewski}, Z.~{Luki{\'c}} and E.~{Van Andel}, \emph{{Nyx: A Massively Parallel AMR Code for Computational Cosmology}}, \href{https://doi.org/10.1088/0004-637X/765/1/39}{\emph{\apj} {\bfseries 765} (2013) 39} [\href{https://arxiv.org/abs/1301.4498}{{\ttfamily 1301.4498}}].

\bibitem{Sexton2021}
J.~Sexton, Z.~Lukic, A.~Almgren, C.~Daley, B.~Friesen, A.~Myers et~al., \emph{Nyx: A massively parallel amr code for computational cosmology}, \href{https://doi.org/10.21105/joss.03068}{\emph{Journal of Open Source Software} {\bfseries 6} (2021) 3068}.

\bibitem{Lukic:2015}
Z.~{Luki{\'c}}, C.~W. {Stark}, P.~{Nugent}, M.~{White}, A.~A. {Meiksin} and A.~{Almgren}, \emph{{The Lyman {\ensuremath{\alpha}} forest in optically thin hydrodynamical simulations}}, \href{https://doi.org/10.1093/mnras/stu2377}{\emph{\mnras} {\bfseries 446} (2015) 3697} [\href{https://arxiv.org/abs/1406.6361}{{\ttfamily 1406.6361}}].

\bibitem{Nishimichi:2020tvu}
T.~Nishimichi, G.~D'Amico, M.~M. Ivanov, L.~Senatore, M.~Simonovi\'c, M.~Takada et~al., \emph{{Blinded challenge for precision cosmology with large-scale structure: results from effective field theory for the redshift-space galaxy power spectrum}}, \href{https://doi.org/10.1103/PhysRevD.102.123541}{\emph{Phys. Rev. D} {\bfseries 102} (2020) 123541} [\href{https://arxiv.org/abs/2003.08277}{{\ttfamily 2003.08277}}].

\bibitem{2009JCAP...08..020M}
P.~{McDonald} and A.~{Roy}, \emph{{Clustering of dark matter tracers: generalizing bias for the coming era of precision LSS}}, \href{https://doi.org/10.1088/1475-7516/2009/08/020}{\emph{\jcap} {\bfseries 8} (2009) 20} [\href{https://arxiv.org/abs/0902.0991}{{\ttfamily 0902.0991}}].

\bibitem{Blas:2015qsi}
D.~Blas, M.~Garny, M.~M. Ivanov and S.~Sibiryakov, \emph{{Time-Sliced Perturbation Theory for Large Scale Structure I: General Formalism}}, \href{https://doi.org/10.1088/1475-7516/2016/07/052}{\emph{JCAP} {\bfseries 1607} (2016) 052} [\href{https://arxiv.org/abs/1512.05807}{{\ttfamily 1512.05807}}].

\bibitem{Blas:2016sfa}
D.~Blas, M.~Garny, M.~M. Ivanov and S.~Sibiryakov, \emph{{Time-Sliced Perturbation Theory II: Baryon Acoustic Oscillations and Infrared Resummation}}, \href{https://doi.org/10.1088/1475-7516/2016/07/028}{\emph{JCAP} {\bfseries 1607} (2016) 028} [\href{https://arxiv.org/abs/1605.02149}{{\ttfamily 1605.02149}}].

\bibitem{Ivanov:2018gjr}
M.~M. Ivanov and S.~Sibiryakov, \emph{{Infrared Resummation for Biased Tracers in Redshift Space}}, \href{https://doi.org/10.1088/1475-7516/2018/07/053}{\emph{JCAP} {\bfseries 1807} (2018) 053} [\href{https://arxiv.org/abs/1804.05080}{{\ttfamily 1804.05080}}].

\bibitem{Vasudevan:2019ewf}
A.~Vasudevan, M.~M. Ivanov, S.~Sibiryakov and J.~Lesgourgues, \emph{{Time-sliced perturbation theory with primordial non-Gaussianity and effects of large bulk flows on inflationary oscillating features}}, \href{https://doi.org/10.1088/1475-7516/2019/09/037}{\emph{JCAP} {\bfseries 09} (2019) 037} [\href{https://arxiv.org/abs/1906.08697}{{\ttfamily 1906.08697}}].

\bibitem{Vlah16}
Z.~{Vlah}, U.~{Seljak}, M.~{Yat Chu} and Y.~{Feng}, \emph{{Perturbation theory, effective field theory, and oscillations in the power spectrum}}, \href{https://doi.org/10.1088/1475-7516/2016/03/057}{\emph{\jcap} {\bfseries 2016} (2016) 057} [\href{https://arxiv.org/abs/1509.02120}{{\ttfamily 1509.02120}}].

\bibitem{Chen24}
S.-F. {Chen}, Z.~{Vlah} and M.~{White}, \emph{{The bispectrum in Lagrangian perturbation theory}}, \href{https://doi.org/10.1088/1475-7516/2024/11/012}{\emph{\jcap} {\bfseries 2024} (2024) 012} [\href{https://arxiv.org/abs/2406.00103}{{\ttfamily 2406.00103}}].

\bibitem{McDonald:1999dt}
P.~McDonald, J.~Miralda-Escude, M.~Rauch, W.~L.~W. Sargent, T.~A. Barlow, R.~Cen et~al., \emph{{The Observed probability distribution function, power spectrum, and correlation function of the transmitted flux in the Lyman-alpha forest}}, \href{https://doi.org/10.1086/317079}{\emph{Astrophys. J.} {\bfseries 543} (2000) 1} [\href{https://arxiv.org/abs/astro-ph/9911196}{{\ttfamily astro-ph/9911196}}].

\bibitem{Kaiser:1987qv}
N.~Kaiser, \emph{{Clustering in real space and in redshift space}}, {\emph{Mon. Not. Roy. Astron. Soc.} {\bfseries 227} (1987) 1}.

\bibitem{BOSS:2016wmc}
{\scshape BOSS} collaboration, \emph{{The clustering of galaxies in the completed SDSS-III Baryon Oscillation Spectroscopic Survey: cosmological analysis of the DR12 galaxy sample}}, \href{https://doi.org/10.1093/mnras/stx721}{\emph{Mon. Not. Roy. Astron. Soc.} {\bfseries 470} (2017) 2617} [\href{https://arxiv.org/abs/1607.03155}{{\ttfamily 1607.03155}}].

\bibitem{Irsic:2018hhg}
V.~Ir\v{s}i\v{c} and M.~McQuinn, \emph{{Absorber Model: the Halo-like model for the Lyman-$\alpha$ forest}}, \href{https://doi.org/10.1088/1475-7516/2018/04/026}{\emph{JCAP} {\bfseries 04} (2018) 026} [\href{https://arxiv.org/abs/1801.02671}{{\ttfamily 1801.02671}}].

\bibitem{Philcox:2020zyp}
O.~H.~E. Philcox, M.~M. Ivanov, M.~Zaldarriaga, M.~Simonovic and M.~Schmittfull, \emph{{Fewer Mocks and Less Noise: Reducing the Dimensionality of Cosmological Observables with Subspace Projections}}, \href{https://doi.org/10.1103/PhysRevD.103.043508}{\emph{Phys. Rev. D} {\bfseries 103} (2021) 043508} [\href{https://arxiv.org/abs/2009.03311}{{\ttfamily 2009.03311}}].

\bibitem{Sailer:2024coh}
N.~Sailer et~al., \emph{{Cosmological constraints from the cross-correlation of DESI Luminous Red Galaxies with CMB lensing from Planck PR4 and ACT DR6}},  \href{https://arxiv.org/abs/2407.04607}{{\ttfamily 2407.04607}}.

\bibitem{Brinckmann:2018cvx}
T.~Brinckmann and J.~Lesgourgues, \emph{{MontePython 3: boosted MCMC sampler and other features}}, \href{https://doi.org/10.1016/j.dark.2018.100260}{\emph{Phys. Dark Univ.} {\bfseries 24} (2019) 100260} [\href{https://arxiv.org/abs/1804.07261}{{\ttfamily 1804.07261}}].

\bibitem{Audren:2011ne}
B.~Audren and J.~Lesgourgues, \emph{{Non-linear matter power spectrum from Time Renormalisation Group: efficient computation and comparison with one-loop}}, \href{https://doi.org/10.1088/1475-7516/2011/10/037}{\emph{JCAP} {\bfseries 1110} (2011) 037} [\href{https://arxiv.org/abs/1106.2607}{{\ttfamily 1106.2607}}].

\bibitem{1992StaSc...7..457G}
A.~{Gelman} and D.~B. {Rubin}, \emph{{Inference from Iterative Simulation Using Multiple Sequences}}, \href{https://doi.org/10.1214/ss/1177011136}{\emph{Statistical Science} {\bfseries 7} (1992) 457}.

\bibitem{Lewis:2019xzd}
A.~Lewis, \emph{{GetDist: a Python package for analysing Monte Carlo samples}},  \href{https://arxiv.org/abs/1910.13970}{{\ttfamily 1910.13970}}.

\bibitem{Diego_Blas_2011}
D.~Blas, J.~Lesgourgues and T.~Tram, \emph{The cosmic linear anisotropy solving system (class). part ii: Approximation schemes}, \href{https://doi.org/10.1088/1475-7516/2011/07/034}{\emph{Journal of Cosmology and Astroparticle Physics} {\bfseries 2011} (2011) 034–034}.

\bibitem{Chudaykin:2020aoj}
A.~Chudaykin, M.~M. Ivanov, O.~H.~E. Philcox and M.~Simonovi\'c, \emph{{Nonlinear perturbation theory extension of the Boltzmann code CLASS}}, \href{https://doi.org/10.1103/PhysRevD.102.063533}{\emph{Phys. Rev. D} {\bfseries 102} (2020) 063533} [\href{https://arxiv.org/abs/2004.10607}{{\ttfamily 2004.10607}}].

\bibitem{SDSS-BAO}
S.~{Alam}, M.~{Ata}, S.~{Bailey}, F.~{Beutler}, D.~{Bizyaev}, J.~A. {Blazek} et~al., \emph{{The clustering of galaxies in the completed SDSS-III Baryon Oscillation Spectroscopic Survey: cosmological analysis of the DR12 galaxy sample}}, {\emph{ArXiv e-prints} (2016) } [\href{https://arxiv.org/abs/1607.03155}{{\ttfamily 1607.03155}}].

\bibitem{Alcock:1979mp}
C.~Alcock and B.~Paczynski, \emph{{An evolution free test for non-zero cosmological constant}}, \href{https://doi.org/10.1038/281358a0}{\emph{Nature} {\bfseries 281} (1979) 358}.

\bibitem{Crocce:2007dt}
M.~Crocce and R.~Scoccimarro, \emph{{Nonlinear Evolution of Baryon Acoustic Oscillations}}, \href{https://doi.org/10.1103/PhysRevD.77.023533}{\emph{Phys. Rev.} {\bfseries D77} (2008) 023533} [\href{https://arxiv.org/abs/0704.2783}{{\ttfamily 0704.2783}}].

\bibitem{Eisenstein:2006nj}
D.~J. Eisenstein, H.-j. Seo and M.~J. White, \emph{{On the Robustness of the Acoustic Scale in the Low-Redshift Clustering of Matter}}, \href{https://doi.org/10.1086/518755}{\emph{Astrophys. J.} {\bfseries 664} (2007) 660} [\href{https://arxiv.org/abs/astro-ph/0604361}{{\ttfamily astro-ph/0604361}}].

\bibitem{2009PhRvD..80f3508P}
N.~{Padmanabhan} and M.~{White}, \emph{{Calibrating the baryon oscillation ruler for matter and halos}}, \href{https://doi.org/10.1103/PhysRevD.80.063508}{\emph{\prd} {\bfseries 80} (2009) 063508} [\href{https://arxiv.org/abs/0906.1198}{{\ttfamily 0906.1198}}].

\bibitem{Sherwin:2012_BAO}
B.~D. {Sherwin} and M.~{Zaldarriaga}, \emph{{Shift of the baryon acoustic oscillation scale: A simple physical picture}}, \href{https://doi.org/10.1103/PhysRevD.85.103523}{\emph{\prd} {\bfseries 85} (2012) 103523} [\href{https://arxiv.org/abs/1202.3998}{{\ttfamily 1202.3998}}].

\bibitem{McQuinn:2015tva}
M.~McQuinn and M.~White, \emph{{Cosmological perturbation theory in 1+1 dimensions}}, \href{https://doi.org/10.1088/1475-7516/2016/01/043}{\emph{JCAP} {\bfseries 01} (2016) 043} [\href{https://arxiv.org/abs/1502.07389}{{\ttfamily 1502.07389}}].

\bibitem{Mcquinn2011}
M.~McQuinn and M.~White, \emph{On estimating ly $\alpha$ forest correlations between multiple sightlines}, {\emph{Monthly Notices of the Royal Astronomical Society} {\bfseries 415} (2011) 2257}.

\bibitem{Simon:2022csv}
T.~Simon, P.~Zhang and V.~Poulin, \emph{{Cosmological inference from the EFTofLSS: the eBOSS QSO full-shape analysis}}, \href{https://doi.org/10.1088/1475-7516/2023/07/041}{\emph{JCAP} {\bfseries 07} (2023) 041} [\href{https://arxiv.org/abs/2210.14931}{{\ttfamily 2210.14931}}].

\bibitem{eBOSS:2017ozs}
{\scshape eBOSS} collaboration, \emph{{Clustering of quasars in SDSS-IV eBOSS : study of potential systematics and bias determination}}, \href{https://doi.org/10.1088/1475-7516/2017/07/017}{\emph{JCAP} {\bfseries 07} (2017) 017} [\href{https://arxiv.org/abs/1705.04718}{{\ttfamily 1705.04718}}].

\bibitem{Schmittfull:2018yuk}
M.~Schmittfull, M.~Simonović, V.~Assassi and M.~Zaldarriaga, \emph{{Modeling Biased Tracers at the Field Level}}, \href{https://doi.org/10.1103/PhysRevD.100.043514}{\emph{Phys.\ Rev.\ D} {\bfseries 100} (2019) 043514} [\href{https://arxiv.org/abs/1811.10640}{{\ttfamily 1811.10640}}].

\bibitem{Schmittfull:2020trd}
M.~Schmittfull, M.~Simonovi\'c, M.~M. Ivanov, O.~H.~E. Philcox and M.~Zaldarriaga, \emph{{Modeling Galaxies in Redshift Space at the Field Level}}, \href{https://doi.org/10.1088/1475-7516/2021/05/059}{\emph{JCAP} {\bfseries 05} (2021) 059} [\href{https://arxiv.org/abs/2012.03334}{{\ttfamily 2012.03334}}].

\bibitem{Ivanov:2024hgq}
M.~M. Ivanov, C.~Cuesta-Lazaro, S.~Mishra-Sharma, A.~Obuljen and M.~W. Toomey, \emph{{Full-shape analysis with simulation-based priors: Constraints on single field inflation from BOSS}}, \href{https://doi.org/10.1103/PhysRevD.110.063538}{\emph{Phys. Rev. D} {\bfseries 110} (2024) 063538} [\href{https://arxiv.org/abs/2402.13310}{{\ttfamily 2402.13310}}].

\bibitem{Ivanov:2024xgb}
M.~M. Ivanov, A.~Obuljen, C.~Cuesta-Lazaro and M.~W. Toomey, \emph{{Full-shape analysis with simulation-based priors: cosmological parameters and the structure growth anomaly}},  \href{https://arxiv.org/abs/2409.10609}{{\ttfamily 2409.10609}}.

\bibitem{Ivanov:2024dgv}
M.~M. Ivanov et~al., \emph{{The Millennium and Astrid galaxies in effective field theory: comparison with galaxy-halo connection models at the field level}},  \href{https://arxiv.org/abs/2412.01888}{{\ttfamily 2412.01888}}.

\bibitem{Bolton17}
J.~S. {Bolton}, E.~{Puchwein}, D.~{Sijacki}, M.~G. {Haehnelt}, T.-S. {Kim}, A.~{Meiksin} et~al., \emph{{The Sherwood simulation suite: overview and data comparisons with the Lyman {$\alpha$} forest at redshifts 2 < z < 5}}, \href{https://doi.org/10.1093/mnras/stw2397}{\emph{\mnras} {\bfseries 464} (2017) 897} [\href{https://arxiv.org/abs/1605.03462}{{\ttfamily 1605.03462}}].

\bibitem{Seljak:2012tp}
U.~Seljak, \emph{{Bias, redshift space distortions and primordial nongaussianity of nonlinear transformations: application to Lyman alpha forest}}, \href{https://doi.org/10.1088/1475-7516/2012/03/004}{\emph{JCAP} {\bfseries 03} (2012) 004} [\href{https://arxiv.org/abs/1201.0594}{{\ttfamily 1201.0594}}].

\bibitem{Bernardeau:2001CPT}
F.~Bernardeau, S.~Colombi, E.~Gaztanaga and R.~Scoccimarro, \emph{{Large scale structure of the universe and cosmological perturbation theory}}, \href{https://doi.org/10.1016/S0370-1573(02)00135-7}{\emph{\physrep} {\bfseries 367} (2002) 1} [\href{https://arxiv.org/abs/astro-ph/0112551}{{\ttfamily astro-ph/0112551}}].

\bibitem{White:2014gfa}
M.~White, \emph{{The Zel'dovich approximation}}, \href{https://doi.org/10.1093/mnras/stu209}{\emph{Mon. Not. Roy. Astron. Soc.} {\bfseries 439} (2014) 3630} [\href{https://arxiv.org/abs/1401.5466}{{\ttfamily 1401.5466}}].

\bibitem{1974ApJ...187..425P}
W.~H. {Press} and P.~{Schechter}, \emph{{Formation of Galaxies and Clusters of Galaxies by Self-Similar Gravitational Condensation}}, \href{https://doi.org/10.1086/152650}{\emph{\apj} {\bfseries 187} (1974) 425}.

\bibitem{1999MNRAS.308..119S}
R.~K. {Sheth} and G.~{Tormen}, \emph{{Large-scale bias and the peak background split}}, \href{https://doi.org/10.1046/j.1365-8711.1999.02692.x}{\emph{\mnras} {\bfseries 308} (1999) 119} [\href{https://arxiv.org/abs/astro-ph/9901122}{{\ttfamily astro-ph/9901122}}].

\bibitem{schlegel_megamapper_concept}
D.~J. {Schlegel}, J.~A. {Kollmeier}, G.~{Aldering}, S.~{Bailey}, C.~{Baltay}, C.~{Bebek} et~al., \emph{{The MegaMapper: A Stage-5 Spectroscopic Instrument Concept for the Study of Inflation and Dark Energy}}, \href{https://doi.org/10.48550/arXiv.2209.04322}{\emph{arXiv e-prints} (2022) arXiv:2209.04322} [\href{https://arxiv.org/abs/2209.04322}{{\ttfamily 2209.04322}}].

\bibitem{DESI:2016}
{DESI Collaboration}, A.~{Aghamousa}, J.~{Aguilar}, S.~{Ahlen}, S.~{Alam}, L.~E. {Allen} et~al., \emph{{The DESI Experiment Part I: Science,Targeting, and Survey Design}}, {\emph{arXiv e-prints} (2016) arXiv:1611.00036} [\href{https://arxiv.org/abs/1611.00036}{{\ttfamily 1611.00036}}].

\bibitem{2016sf2a.conf..259P}
M.~M. {Pieri}, S.~{Bonoli}, J.~{Chaves-Montero}, I.~{P{\^a}ris}, M.~{Fumagalli}, J.~S. {Bolton} et~al., \emph{{WEAVE-QSO: A Massive Intergalactic Medium Survey for the William Herschel Telescope}},  in \emph{SF2A-2016: Proceedings of the Annual meeting of the French Society of Astronomy and Astrophysics}, C.~{Reyl{\'e}}, J.~{Richard}, L.~{Cambr{\'e}sy}, M.~{Deleuil}, E.~{P{\'e}contal}, L.~{Tresse} et~al., eds., pp.~259--266, Dec., 2016, \href{https://arxiv.org/abs/1611.09388}{{\ttfamily 1611.09388}}, \href{https://doi.org/10.48550/arXiv.1611.09388}{DOI}.

\bibitem{2022PFSGE}
J.~{Greene}, R.~{Bezanson}, M.~{Ouchi}, J.~{Silverman} and {the PFS Galaxy Evolution Working Group}, \emph{{The Prime Focus Spectrograph Galaxy Evolution Survey}}, \href{https://doi.org/10.48550/arXiv.2206.14908}{\emph{arXiv e-prints} (2022) arXiv:2206.14908} [\href{https://arxiv.org/abs/2206.14908}{{\ttfamily 2206.14908}}].

\bibitem{2019Msngr.175....3D}
R.~S. {de Jong}, O.~{Agertz}, A.~A. {Berbel}, J.~{Aird}, D.~A. {Alexander}, A.~{Amarsi} et~al., \emph{{4MOST: Project overview and information for the First Call for Proposals}}, \href{https://doi.org/10.18727/0722-6691/5117}{\emph{The Messenger} {\bfseries 175} (2019) 3} [\href{https://arxiv.org/abs/1903.02464}{{\ttfamily 1903.02464}}].

\end{thebibliography}\endgroup

\end{document}